\documentclass[aps,pre,preprint,showpacs]{revtex4}

\usepackage{graphicx}
\usepackage{lipsum}
\usepackage{float}
\usepackage{amsmath}
\usepackage{amssymb}
\usepackage{mathtools}
\include{mathmacros}

\begin{document}

\title{
The distribution of shortest path lengths 
in a class of node duplication network models
}

\author{Chanania Steinbock}

\affiliation{
Racah Institute of Physics, 
The Hebrew University, 
Jerusalem 91904, Israel}

\author{Ofer Biham} 

\affiliation{
Racah Institute of Physics, 
The Hebrew University, 
Jerusalem 91904, Israel}

\author{Eytan Katzav} 

\affiliation{
Racah Institute of Physics, 
The Hebrew University, 
Jerusalem 91904, Israel}

\begin{abstract}
We present analytical results for the 
distribution of shortest path lengths (DSPL)
in a network growth model which evolves by node duplication (ND).
The model captures essential properties of the structure and growth
dynamics of social networks,
acquaintance networks 
and scientific citation networks,
where duplication mechanisms play a major role.
Starting from an initial seed network, at each time step
a random node, referred to as a mother node, is selected for duplication.
Its daughter node is added to the network, forming a link to the 
mother node, and with probability $p$ to each one of its neighbors.
The degree distribution of the resulting network turns out to
follow a power-law distribution, 
thus the ND network 
is a scale-free network.
To calculate the DSPL we
derive a master equation for the time
evolution of the probability 
$P_t(L=\ell)$, $\ell=1,2,\dots$, 
where 
$L$ is the distance between a pair of nodes and $t$ is the time.
Finding an exact analytical solution of the master equation, 
we obtain a closed form expression for
$P_t(L=\ell)$.
The mean distance, 
$\langle L \rangle_t$, 
and the diameter,
$\Delta_t$,
are found to scale like
$\ln t$,
namely the ND network is a small world network.
The variance of the DSPL is also found to scale like 
$\ln t$.
Interestingly, the mean distance
and the diameter exhibit properties of a small world network, 
rather than the ultrasmall world network behavior
observed in other scale-free networks, 
in which 
$\langle L \rangle_t \sim \ln \ln t$.

\end{abstract}

\maketitle

\section{Introduction}

The increasing interest in the field of 
complex networks in recent years is motivated by 
the realization that a large variety of systems and processes
in physics, chemistry, biology, engineering, and society 
can be usefully described by network models
\cite{Albert2002,Caldarelli2007,Havlin2010,Newman2010,Estrada2011b,Barrat2012}.
These models consist of nodes and edges, where the nodes
represent physical objects, while the edges represent the
interactions between them.
It was found that networks appearing in different contexts often
share various structural properties.
For example, they exhibit repeating network motifs
such as the feed-forward loop (FFL) and the auto-regulator
\cite{Milo2002,Alon2006}.
The structure of these motifs and their abundance
provide useful information on the growth mechanism
of the network and often has functional importance.
At the global scale,
many of these networks are scale-free,
which means that they exhibit power-law degree distributions of the form 
$P(K=k) \sim k^{-\gamma}$
\cite{Barabasi1999,Jeong2000,Krapivsky2000,Krapivsky2001,Vazquez2003}.
The most highly connected nodes, called hubs, 
play a dominant role in dynamical processes on these networks.
A central feature of random networks is the small-world property,
namely the fact that the mean distance and the diameter 
scale like $\ln N$,
where $N$ is the network size
\cite{Milgram1967,Watts1998,Chung2002,Chung2003}.
Moreover, it was shown that scale-free networks are generically
ultrasmall, namely their mean distance and diameter scale like
$\ln \ln N$
\cite{Cohen2003}.

While pairs of adjacent nodes exhibit direct interactions,
the interactions between most pairs of nodes
are indirect, and are mediated by intermediate nodes and edges.
Pairs of nodes may be connected by many different paths. 
The shortest among these paths are of particular 
importance because they are likely to provide the fastest 
and strongest interactions.
Therefore, it is of much interest to study the 
distribution of shortest path lengths (DSPL) 
between pairs of nodes in different types of networks.
Such distributions, 
which are also referred to as distance distributions,
are expected to depend on the network 
structure and size.
They are of great importance for the 
temporal evolution of dynamical processes
\cite{Barrat2012} 
such as signal propagation in genetic regulatory networks
\cite{Giot2003,Maayan2005}, 
navigation 
\cite{Dijkstra1959,Delling2009}
and epidemic spreading 
\cite{Satorras2015}.
Central measures of the DSPL such as 
the mean distance 
and extremal measures such as
the diameter 
were studied
\cite{Bollobas2001,Durrett2007,Watts1998,Fronczak2004,Newman2001b}.
However, apart from a few studies
\cite{Newman2001,Dorogotsev2003,Blondel2007,Hofstad2007,Esker2008,Shao2008,Shao2009},
the DSPL has not attracted nearly as much 
attention as the degree distribution.
Recently, an analytical approach was developed for calculating 
the DSPL 
\cite{Katzav2015}
in the
Erd{\H o}s-R\'enyi (ER) network
\cite{Erdos1959},
which is the simplest mathematical model of a random network.
More general formulations were later developed 
\cite{Nitzan2016,Melnik2016},
for the broader class of 
configuration model networks
\cite{Molloy1995,Newman2001}. 

To gain insight into the structure of complex networks, 
it is useful to study the growth dynamics that gives rise to these structures.
In general, it appears that many of the networks encountered in biological,
ecological and social systems grow step by step, by the addition of new nodes
and their attachment to existing nodes. In some networks, the new nodes 
emerge with no predefined connections, while in other networks
the new nodes result from the duplication of existing nodes, 
followed by a stochastic readjustment of their links.
A fundamental feature of these growth processes is
the preferential attachment mechanism, in which the likelihood of
an existing node to gain a link to the new node 
is proportional to its degree.
It was shown that growth models based on preferential
attachment give rise to scale-free networks,
which exhibit power-law degree distributions
\cite{Barabasi1999,Albert2002}. 

The effect of node duplication (ND) processes 
on the structure and evolution of networks
was studied using a simple network growth model. 
In this model,  
at each time step a random node, referred to as a mother node,
is selected for duplication
\cite{Bhan2002,Kim2002,Chung2003b,Krapivsky2005,Ispolatov2005,Ispolatov2005b,Bebek2006,Li2013}.
The new, daughter node, retains a copy of 
each link of the mother node with probability $p$.
In this model the daughter node does not form a link
to the mother node, and thus in the following 
is referred to as the
uncorded ND model. 
In the case that none of these links were retained,
the daughter node remains isolated and
is removed from the network.
In such case, a new mother node is randomly selected for duplication
and the growth process continues.
Note that as $p$ is decreased, the probability that the daughter node will be 
discarded increases and the network growth process slows down.
It was shown that for $0 < p < 1/2$ 
the resulting network exhibits a power law
degree distribution of the form 

\begin{equation}
P_t(K=k) \sim k^{-\gamma}.
\label{eq:powerlaw} 
\end{equation}

\noindent
For 
$0 < p < 1/e$,
where $e$ is the base of the natural logarithm,
the exponent is given by the nontrivial solution of
the equation

\begin{equation}
\gamma = 3 - p^{\gamma-2}, 
\label{eq:gamma}
\end{equation}

\noindent
while for 
$1/e \le p < 1/2$ 
it takes the value 
$\gamma=2$
\cite{Ispolatov2005}.
In the former case the mean degree,
$\langle K \rangle_t$,
converges to 
an asymptotic value
while in the latter case it diverges logarithmically with
the network size.
For $1/2 \le p \le 1$ the degree distribution does not converge at all.

\begin{figure}
\includegraphics[width=10cm]{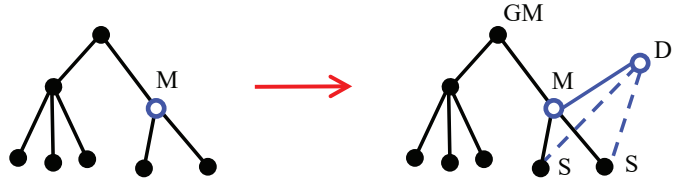}
\caption{
(Color online)
Illustration of the corded ND model.
A random node, referred to as a mother node, M (empty circle)
is selected for duplication. The newly created daughter node, D
(empty circle) forms a deterministic edge (solid line) 
to the mother node, and with probability $p$
it forms a probabilistic edge (dashed line) to each one of the neighbors of M.
In this example, D forms links to its two sister nodes, denoted by S,
but does not form a link to its grand-mother node, denoted by GM.
In this illustration, all the other edges (solid lines) are deterministic edges.
}
\label{fig:1}
\end{figure}

Recently, a new node duplication model
was introduced and studied
\cite{Lambiotte2016,Bhat2016}.
In this model, referred to as the corded ND model, 
starting from a seed network which 
consists of a single connected component
of $s$ nodes,
at each time step a random, mother node, M,
is selected for duplication.
The daughter node, D, is added to the network.
It forms a link to its mother node, M,
and is also connected with probability $p$ to each 
neighbor of M
(Fig. \ref{fig:1}).
\noindent
It was shown that for 
$0 < p < 1/2$ 
the corded ND model generates a sparse network,
while for
$1/2 \le p \le 1$ the model gives rise to a dense network in 
which the mean degree increases with the network size
\cite{Lambiotte2016,Bhat2016}.
The ND models exhibit 
the preferential attachment property.
This is due to the fact that the probability of a node of 
degree $k$ to be a neighbor of the randomly selected mother node is
proportional to $k$.
Therefore, the degrees of the neighbors of the mother node selected at
time $t$ are drawn from the distribution

\begin{equation}
\tilde P_t(K=k) = \frac{k P_t(K=k)}{\langle K \rangle_t}.
\end{equation}

\noindent
The daughter node forms a link to each one of these nodes with probability $p$.
Thus, the probability that the daughter node will form a link to a node of
degree $k$ is proportional to $\tilde P_t(K=k)$.
The degree distribution of the corded ND network was studied
in Refs.
\cite{Lambiotte2016,Bhat2016}.
It was found that for $0 < p < 1/2$,
in the asymptotic limit,
the degree distribution 
of this network follows Eq.
(\ref{eq:powerlaw}),
where the exponent 
$\gamma = \gamma(p)$
is given by the non-trivial solution of the
equation 

\begin{equation}
\gamma = 1 + p^{-1} - p^{\gamma-2}.
\label{eq:gamma2}
\end{equation}

\noindent
This solution, $\gamma=\gamma(p)$
is a monotonically decreasing function of $p$,
in the range of $0 < p < 1/2$.
In the limit of $p \rightarrow 0$, the exponent $\gamma$ diverges like
$\gamma(p) \sim 1/p$, while
$\gamma(1/2)=2$.
In the asymptotic limit,
the mean degree is given by

\begin{equation}
\langle K \rangle = \frac{2}{1-2p},
\label{eq:Kmean}
\end{equation}

\noindent
while the
second moment of the degree distribution is 
given by
\cite{Bhat2016}

\begin{equation}
\langle K^2 \rangle = \left( \frac{2}{1-2p} \right)
\left( \frac{3+2p-p^2}{1-2p-p^2} \right),
\ \ \ \ \ 0 < p < \sqrt{2}-1.
\label{eq:K2mean}
\end{equation}

\noindent
The sparse network regime can be divided into two parts.
For 
$0 < p < \sqrt{2}-1$
the exponent 
$\gamma(p) > 3$,
thus in this range the first two moments,
$\langle K \rangle$ and $\langle K^2 \rangle$,
are finite.
For 
$\sqrt{2} -1 < p < 1/2$,
the exponent $\gamma$ takes
values in the range
$2 < \gamma(p) < 3$,
thus in this range the first moment
is finite while the second moment diverges.
Using Eqs.
(\ref{eq:Kmean})
and
(\ref{eq:K2mean}),
it is found that the connective constant

\begin{equation}
\lambda = \frac{\langle K^2 \rangle - \langle K \rangle}{\langle K \rangle}
\end{equation}

\noindent
of the corded ND network is given by

\begin{equation}
\lambda = \frac{ 2 (1+2p) }{ 1-2p-p^2 }.
\end{equation}

\noindent
While in the asymptotic limit the probability
$P(K=k)$ may be non-zero for any integer value of $k$,
for a finite network of $N$ nodes, it is bounded in the 
range $1 \le k \le k_{\rm max}$,
where $k_{\rm max} = N-1$.

Node duplication processes capture essential features of empirical networks.
For example, an important evolutionary process in genetic regulatory networks 
is gene duplication, and subsequent mutations of one of the copies
\cite{Ohno1970,Teichmann2004}.
As a result, the mutated gene may lose some of its links, and eventually
may also form new links.
Typically, there is no link between the two copies of the duplicated gene
\cite{AutoReg}.
Therefore, the node duplication process 
resembles the uncorded ND model studied 
in Refs.
\cite{Bhan2002,Kim2002,Chung2003b,Krapivsky2005,Ispolatov2005,Ispolatov2005b,Bebek2006,Li2013}.
The corded ND model, introduced in Refs.
\cite{Lambiotte2016,Bhat2016},
is suitable for the modeling of acquaintance networks,
in which a newcomer who has a friend in a 
new community becomes acquainted with members of the friend's social group
\cite{Toivonen2009}.
Unlike the uncorded ND model, 
the formation of triadic
closures is built-in to the dynamics of the corded ND model. 
This means that once the
daughter node forms a link to a neighbor of the mother node, it completes
a triangle in which the mother, neighbor and daughter nodes are all connected 
to each other. The formation of triadic closures is an essential property of
the dynamics of social networks where people tend to 
form a connection to a friend of a friend
\cite{Granovetter1973}. 
Therefore, the corded ND model is more suitable 
for the description of social networks than the uncorded ND model.
The corded ND model also describes scientific citation networks
\cite{Redner1998,Redner2005,Radicchi2008},
in which the nodes represent papers, while the links
represent citations.
While acquaintance networks are undirected, citation networks are
directed networks, with links pointing from the later (citing) paper
to the earlier (cited) paper.
It was found that a paper, A, citing an earlier paper, B,
often also cites one or several papers, C$_1$, C$_2$,$\dots$,C$_r$,
which were cited in B
\cite{Peterson2010}.
The resulting network module consists of $r$ triangles, or triadic closures,
which share the AB edge. 
Since the links of this network are pointing backwards, each
one of these triangles can be considered as a 
feed-backward loop (FBL).

The corded ND model 
exhibits a unique structure, which is radically different from
configuration model networks with the same degree distribution.
Unlike the configuration model network
\cite{Molloy1995,Newman2001}, 
which may include small, isolated clusters,
the corded ND network consists of a single connected component.
Therefore, unlike the configuration model,
it does not exhibit a percolation transition.
Also, while the configuration model network exhibits a local
tree-like structure, the ND network includes a large number of
triangles and other short cycles even in the dilute case
of $0 < p < 1/2$
\cite{Lambiotte2016,Bhat2016}.
Interestingly, many empirical networks exhibit a high 
abundance of triangles,
both in undirected networks
\cite{Newman2001b}
and in directed networks,
where most triangles form FFLs, while triangular
feedback loops are rare
\cite{Milo2002,Alon2006}.

In the special case of $p=0$, the corded ND network 
is a tree, which consists
only of the mother-daughter edges.
This tree turns out to form a backbone for the corded 
ND network at $p>0$, and is thus
refereed to as the backbone tree. 
Once a mother node is selected for duplication,
the mother-daughter edge is added deterministically.
Therefore, the edges of the backbone tree are called deterministic edges.
The other edges, which exist only for $p>0$, are called probabilistic edges.
In the limit of $p=0$, where the corded ND network is
a tree, the shortest path between any pair of nodes is unique.
In fact, on a tree structure the shortest path is the {\it only} path
between any pair of nodes. 
Since the path which resides on the
backbone tree consists only of deterministic edges, 
it is referred to as the deterministic path.
For $p>0$ the tree is decorated by probabilistic edges.
These edges may give rise to alternate paths between any pair of nodes, 
in addition to the deterministic path which fully resides on the backbone tree.
An alternate path may consist of probabilistic edges alone, 
or from a combination
of probabilistic and deterministic edges. 
In case that the deterministic path between a pair of nodes
is shorter than all the alternate paths,
it remains the unique shortest path.
When the shortest among the alternate paths between a pair of nodes are of the
same length as the deterministic path, the shortest path becomes degenerate.
Alternate paths may also be shorter than the deterministic path,
in which case they become the shortest paths.

In this paper we present analytical results for the 
DSPL of the corded ND model.
Focusing on the
sparse network regime of $0 < p < 1/2$,
we derive a master equation for the
time evolution of the probabilities
$P_t(L=\ell)$,
where $\ell=1,2,\dots$ is the distance between a pair of nodes
and $t$ is the time.
The derivation of the master equation requires information on
the structure of the backbone tree and on the degeneracies
of the shortest paths.
Solving the master equation we
obtain an expression for
$P_t(L=\ell)$,
which consists of two convolution-like sums.
The first sum emanates from the
DSPL of the seed network,
$P_0(L=\ell)$,
while the second sum involves a discrete 
exponential function.
We calculate the mean distance,
$\langle L \rangle_t$,
and the diameter,
$\Delta_t$,
and show
that in the long-time limit they scale like
$\ln t$,
namely the corded ND network is a small-world network
\cite{Milgram1967,Watts1998,Chung2002,Chung2003}.
Interestingly, this behavior differs from other scale-free networks
which are ultrasmall, namely their mean distance follows
$\langle L \rangle_t \sim \ln \ln t$
\cite{Cohen2003}.

The paper is organized as follows.
In Sec. II we present the corded ND model.
In Sec. III we analyze the backbone tree,
consisting of the mother-daughter edges.
In Sec. IV we consider the degeneracies of 
the shortest paths in the corded ND network.
Using the results of sections III and IV
we derive, in Sec. V,
a master equation for the time evolution of the DSPL
and solve it analytically.
In Sec. VI we study properties of the DSPL. 
The mean distance is studied in Sec. VII, the diameter is evaluated in
Sec. VIII and the variance of the DSPL is obtained in Sec. IX.
The results are discussed in Sec. X and summarized in Sec. XI.

\section{The corded node duplication model}

Consider the corded ND model introduced in Refs. 
\cite{Lambiotte2016,Bhat2016}.
At each time step, a random node,
referred to as the mother node,
is selected for duplication.
The daughter node is added to the network, 
forming a link to the mother node and
with probability $p$ to each neighbor
of the mother node
\cite{Lambiotte2016,Bhat2016}.
The growth process starts from an initial 
seed network of $N_0=s$ nodes.
Thus, the network size
after $t$ time steps is 
$N_t = t + s$.

\begin{figure}
\includegraphics[width=9.0cm]{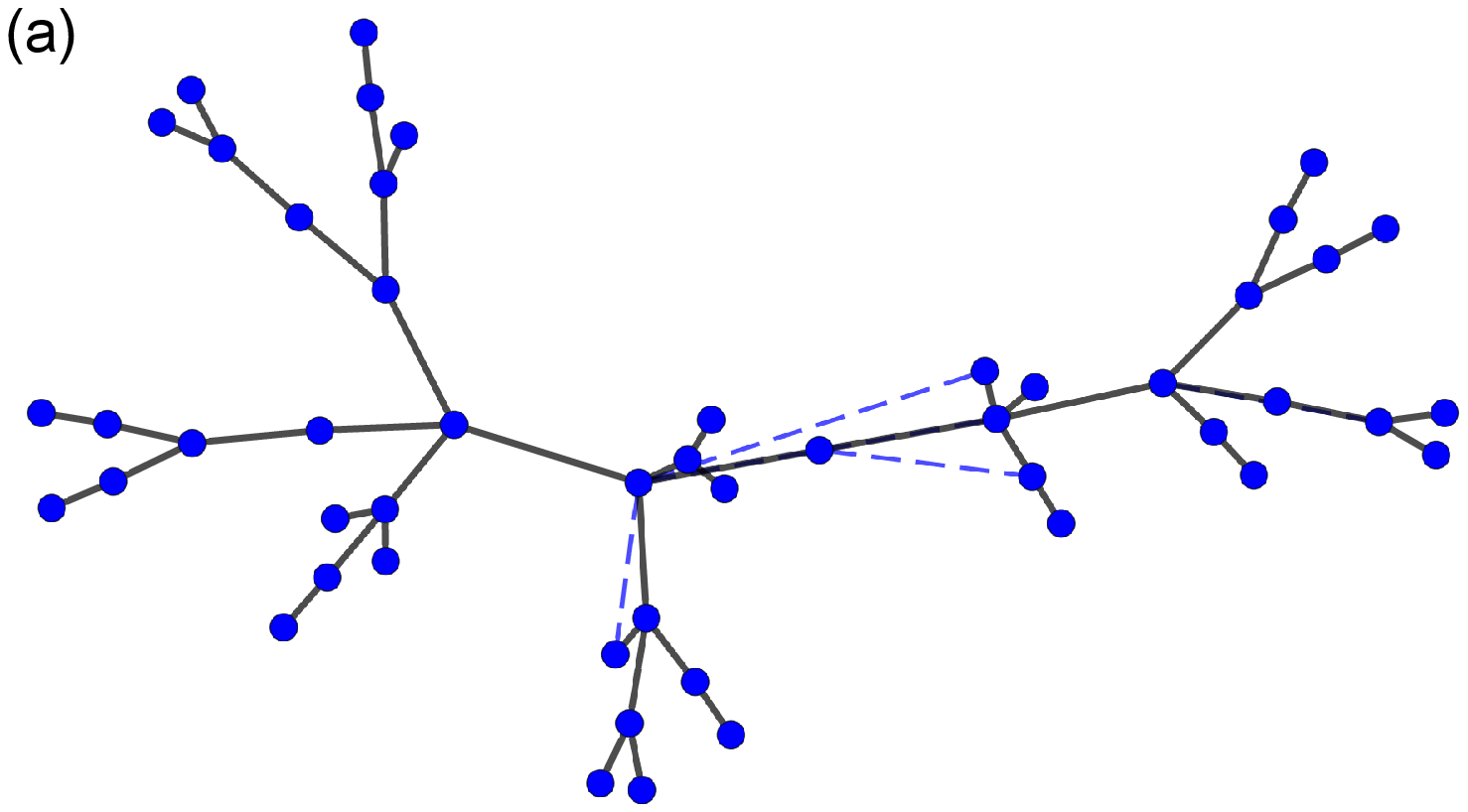} \\
\includegraphics[width=9.0cm]{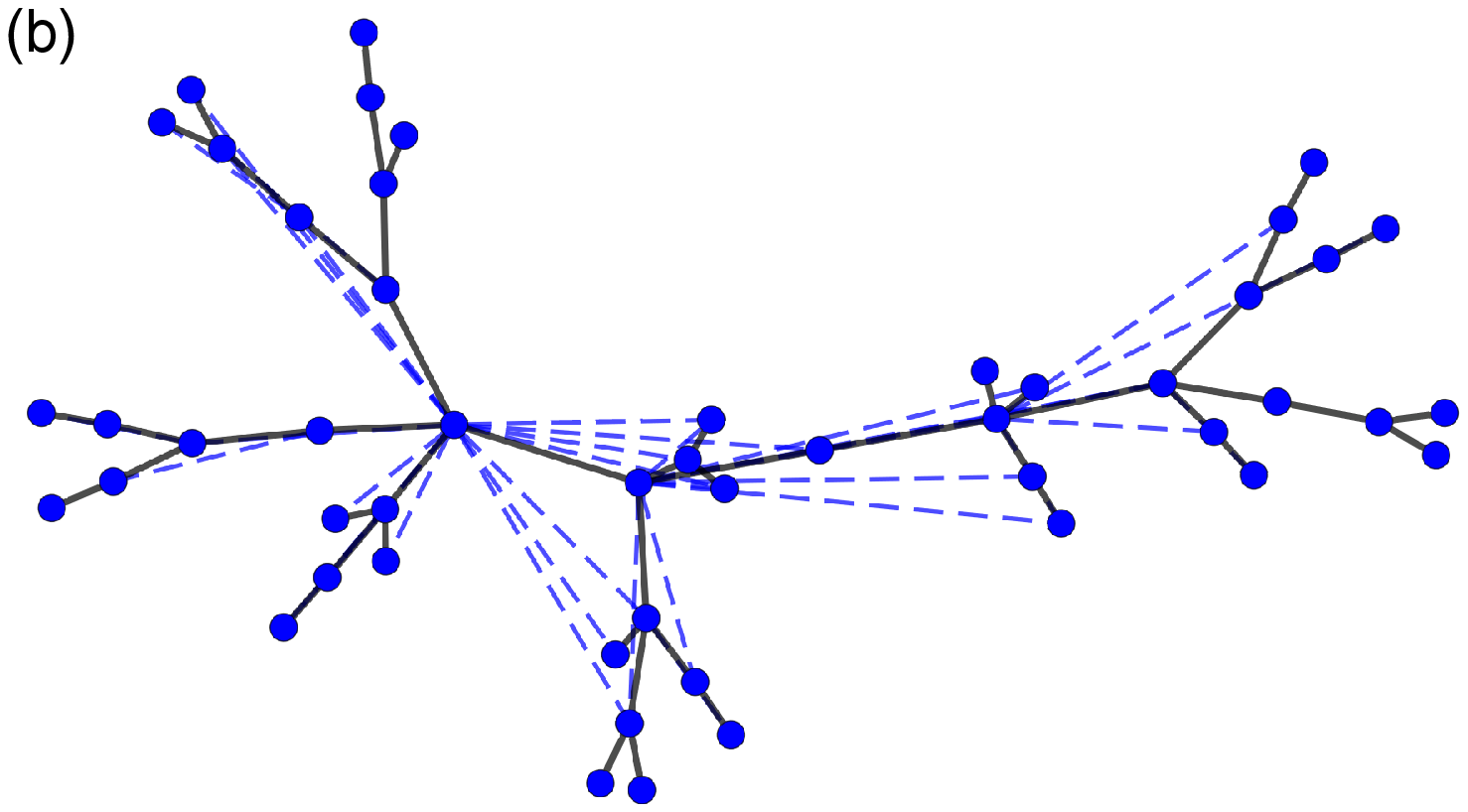}
\caption{
(Color online)
Two instances of corded ND networks of size $N=50$,
with $p=0.1$ (a) and $p=0.4$ (b).
For the sake of comparison, both instances are formed around
the same backbone tree (solid lines).
The probabilistic edges (dashed lines) essentially decorate the
tree. Upon formation, a probabilistic edge shortens the distance 
between its two ends from $L=2$ to $L=1$, forming a triangle.
Increasing $p$ makes the network denser.
}
\label{fig:2}
\end{figure}

In Fig. \ref{fig:2}
we present two instances of the corded ND network, of size $N_t=50$,
which were formed around the same backbone tree.
Both networks were grown from a seed of size $s=2$,
with $p=0.1$ [Fig. \ref{fig:2}(a)] and $p=0.4$ [Fig. \ref{fig:2}(b)].
Each network instance includes $N_t-1=49$ deterministic edges
(solid lines).
The network of Fig. \ref{fig:2}(b) is denser and includes $21$
probabilistic edges (dashed lines), 
compared to $3$ probabilistic edges in Fig. \ref{fig:2}(a).

The corded ND model exhibits many interesting properties.
Since the mother node at time $t$ is selected randomly from
all the $N_t$ nodes in the network, its degree
is effectively drawn from the degree distribution $P_t(K=k)$.
The mother node gains a link to the daughter node, thus
its degree increases by 1.
By construction,
the degree
of the daughter node
cannot exceed the degree of the mother node.
In case that all the links are duplicated, the degree of the
daughter node is equal to the degree of the mother node, 
while in case that none of them
is duplicated the degree of the daughter node is 1.

In order to obtain a connected network, it is required
that the seed network will consist of a single connected
component.
The size of the seed network is denoted by
$s$ and its degree distribution is $P_0(K=k)$.
The mean degree of the
seed network is denoted by
$\langle K \rangle_0$.
The DSPL of the seed network is denoted by
$P_0(L=\ell)$
and the mean distance is denoted by
$\langle L \rangle_0$.
The DSPL and the mean degree are related by
$P_0(L=1) = \langle K \rangle_0 / (s-1)$.
The probability $P_0(L=\ell)$ may take non-zero values
for $\ell=1,2,\dots,\Delta_0$,
where $\Delta_0$ is the diameter of the seed network,
while $P_0(L=\ell)=0$ for $\ell \ge \Delta_0 + 1$.
For seed networks of $s$ nodes, 
$\Delta_0$ may take values in the range
$1 \le \Delta_0 \le s-1$.

The most convenient choice of a seed 
network is a complete graph of $s$ nodes.
In this case, the degree distribution of the seed network is
$P_0(K=k) = \delta_{k,s-1}$.
The DSPL of the seed network is
$P_0(L=\ell) = \delta_{\ell,1}$,
where $\delta_{i,j}$ is the Kronecker delta,
and its diameter is $\Delta_0=1$.
To avoid memory effects, which slow down the
convergence to the asymptotic structure, it is often convenient to use
a seed network which consists of a single node,
namely $s=1$.
In this case
the degree distribution of the seed network is given by
$P_0(K=k) = \delta_{k,0}$,
while its DSPL
is not defined.
However, 
the DSPL becomes well defined
at time $t=1$, when the network consists of
a pair of connected nodes,
whose degree distribution is given by
$P_1(K=k) = \delta_{k,1}$,
its DSPL is
$P_1(L=\ell) = \delta_{\ell,1}$
and its diameter is $\Delta_1=1$.
Another interesting choice for the seed network is a
linear chain of $s$ nodes.
In this case, the 
initial degree distribution is
$P_0(K=k) =  (2/s) \delta_{k,1} + (1-2/s) \delta_{k,2}$,
and the initial DSPL is

\begin{equation}
P_0(L=\ell) = \frac{s-\ell}{
\binom{s}{2}
},
\end{equation} 

\noindent
for $\ell=1,2,\dots,s-1$.
This choice captures the largest possible diameter
in a seed network of $s$ nodes, namely $\Delta_0=s-1$.

\section{The backbone tree}

The mother-daughter links in the
corded ND network form a random tree structure,
which serves as a backbone tree for the resulting network.
The backbone tree is a random recursive tree
\cite{Smythe1995,Drmota1997,Drmota2005}.
To study its properties, one can take the limit of $p=0$,
in which the corded ND network is reduced to the backbone tree.
The degree distribution of the backbone tree,
denoted by
$P^{\rm B}_t(K=k)$,
evolves in time according to

\begin{equation}
P^{\rm B}_{t+1}(K=k) = 
\frac{1}{N_t+1} \left[ (N_t-1)P^{\rm B}_t(K=k) + P^{\rm B}_t(K=k-1) + \delta_{k,1} \right].
\label{eq:recK}
\end{equation}

\noindent
The second term on the right hand side accounts for the degree of the
mother node, which increases by 1 due to the link to the daughter node.
The third term accouts for the degree of the daughter node
(which is $K=1$), while the
first term accounts for all the other nodes in the network.
Subtracting $P^{\rm B}_t(K=k)$ from both sides 
of Eq. (\ref{eq:recK})
and replacing the difference on the left
hand side by a time derivative we obtain

\begin{equation}
\frac{d}{dt} P^{\rm B}_t(K=k) = 
\frac{1}{N_t+1} \left[ - 2 P^{\rm B}_t(K=k) + P^{\rm B}_t(K=k-1)
+ \delta_{k,1} \right].
\label{eq:mera}
\end{equation}

\noindent
In the long time limit, the degree distribution is expected to reach a steady
state, in which the time derivative vanishes. The steady state solution of Eq.
(\ref{eq:mera})
is given by

\begin{equation}
P^{\rm B}(K=k) = \frac{1}{2^k}.
\label{eq:PbK}
\end{equation}

\noindent
The corresponding tail distribution is given by
$P^{\rm B}(K > k) = 1/2^k$.
Note that the degree distribution of the backbone tree,
given by Eq.
(\ref{eq:PbK}),
is a discrete exponential distribution.
It is very different from the degree distribution of the full
corded ND network, which is a power-law distribution.
Eq.
(\ref{eq:PbK})
captures important properties of the network.
In particular, it shows that half of the nodes 
in the backbone tree are leaf nodes,
which have only one link.
One fourth of the nodes in the backbone tree have two links,
namely they lie along linear chains with no branching.
The remaining nodes 
are branching points with three or more links.

It is useful to define a conditional degree distribution
of the form,
$P^{\rm B}(K=k | K > k_0)$,
namely the degree distribution of all the nodes of degree
$K > k_0$.
The conditional degree distribution can be expressed in the form

\begin{equation}
P^{\rm B}(K=k | K > k_0) = \frac{P^{\rm B}(K=k; K>k_0)}{P^{\rm B}(K>k_0)}.
\end{equation}

\noindent
Thus, it is given by

\begin{equation}
P^{\rm B}(K=k|K>k_0) = \frac{1}{2^{k-k_0}}.
\label{eq:Pb_cond}
\end{equation}

\noindent
For example, this means that nodes which are not leaves (namely
of degree $k>1$), are of degree $2$ with probability of $1/2$, are of degree
$3$ with probability of $1/4$, and so on.

\section{The degeneracy of the shortest paths}

Consider a pair of nodes, $i$ and $j$, which are at a distance $L=\ell$
from each other. The shortest path from $i$ to $j$ may be unique or it
may be degenerate. 
In case that the shortest path is degenerate,
there are at least two different paths of
length $\ell$ from $i$ to $j$ (which may have overlapping segments).
In particular, the degenerate paths may differ in the first step, starting
from node $i$.
Here we focus on the degeneracy of the first step, namely on the
number of neighbors of node $i$ which reside on shortest paths
from $i$ to $j$. We denote the distribution of degeneracy levels of
the first steps of the shortest paths by $P(G=g)$, where $g=1,2,\dots$.
In order to calculate the distribution $P(G=g)$ we follow the growth 
process of the network and consider the shortest path from the newly
formed daughter node, D, to a randomly selected target node T. 
It is important to note that the distances $L_{\rm DT}$ between the daughter node, D, and
all the existing nodes, T, in the network are determined upon formation
of the node D. This is due to the fact that nodes and edges which will be added later cannot 
form paths between D and T which are shorter than $L_{\rm DT}$.
However, they can form additional paths of length $L_{\rm DT}$,
thus increasing the degeneracy of the shortest paths.

Since the shortest paths on the backbone tree are unique, it is expected
that for $p \ll 1/2$ the shortest paths between most pairs of nodes will
not be degenerate. Moreover, it is expected that degenerate paths will
exhibit low degeneracy level, namely the probability $P(G=g)$ will sharply
decrease as $g$ is increased. Therefore, we will focus below on the 
probability of a double degeneracy, $P(G=2)$.

It turns out that there are two growth scenarios which give rise to
a double degeneracy of the shortest path from the daughter node,
D, to a random target node T.
In the first scenario, two probabilistic edges form an
alternate path of length $L=2$ between nodes D and GM,
which is degenerate with the shortest
path which goes along the branch of the backbone tree.
In the second scenario, there are two probabilistic edges which
form shortcuts between pairs of nodes which are next nearest
neighbors on the backbone tree. As a result, they give rise to
two degenerate paths of length $L=2$, where each path consists
of one deterministic edge and one probabilistic edge.

\begin{figure}
\includegraphics[width=9cm]{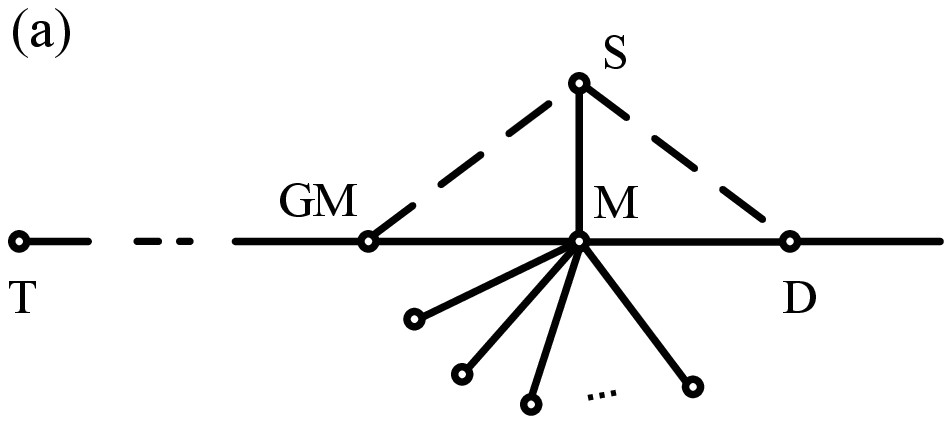} \\
\vspace{0.2in}
\includegraphics[width=9cm]{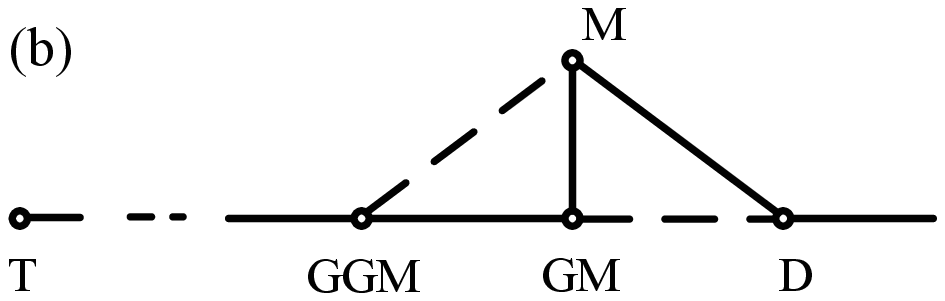}
\caption{
Illustrations of two local network structures which 
give rise to double degeneracy,
$G=2$, in the first step of the shortest 
path between the daughter node, D,
on the right and a target node, T, which resides 
further down the branch, on the left. 
In both illustrations, solid lines correspond to deterministic edges,
which belongs to 
the backbone tree, while dashed lines correspond to probabilistic edges.
(a) An alternate path of length $L=2$ is formed by probabilistic edges
between nodes D and GM, via a sister node, S. This path is degenerate
with the primary path which resides fully on the backbone tree.
Such structure may form in two distinct sequences of events. 
One possibility is that S is an older sister which was connected to GM
upon formation. When D forms it connects probabilistically to S 
and completes the alternate path.
In the other possibility, S is a younger sister of D,
conditioned on D not forming a probabilistic edge to GM
upon formation. When S is formed, it connects simultaneously
to GM and to D, thus forming the alternate path.
(b) In this structure, the distance between D and GGM along the
backbone tree is $L=3$, while the shortest paths in the entire network
is of length $L=2$. This is achieved by two consecutive probabilistic 
shortcuts, one from M to GGM (created upon formation of M)
and the other from D to GM (created upon formation of D).
Note that in this case, the existence of sisters of D (younger or older)
makes no difference.
}
\label{fig:3}
\end{figure}

The first scenario is shown in Fig. \ref{fig:3}(a).
In this scenario, the node D has an older sister, S, 
which is connected to node GM via a probabilistic edge.
In case that D forms a probabilistic edge to S, these
two probabilistic edges form an alternate path of length 
$L=2$ from D to GM. The probability of this scenario is proportional to $p^2$.
In general, node D may have several sister nodes.
The number of such sister nodes is given by $k-2$, where
$k$ is the degree of the mother node, M.
Therefore, the probability that the path from D to T will be doubly
degenerate due to the mechanism of Fig. 
\ref{fig:3}(a) is

\begin{equation}
P(G=2) = \sum_{k=3}^{\infty} {
\binom{k-2}{1}
} P^{\rm B}(K=k | K>2) p^2
(1-p^2)^{k-3},
\label{eq:g2}
\end{equation}

\noindent
where
$P(K=k | K>2)$
is the conditional degree distribution of the backbone tree, given by Eq.
(\ref{eq:Pb_cond}).
Evaluating the right hand side of Eq.
(\ref{eq:g2})
we find that 
$P(G=2)=p^2 + O(p^4)$.

The second scenario is shown in 
Fig. \ref{fig:3}(b).
In this case, the mother node, M, is connected not only to its own
mother node, GM, but also 
(with probability $p$)
to its grandmother node, referred to as GGM. 
Upon formation of node D, it may form (with probability $p$)
a probabilistic edge to node GM. 
In such case, there are two degenerate paths from D to GGM.
The probability of this scenario is 
$P(G=g) = p^2 + O(p^4)$.

It can be shown that the two scenarios presented above are mutually exclusive,
thus the overall probability for the shortest path to be doubly
degenerate is 

\begin{equation}
P(G=2) = P_a(G=2) + P_b(G=2) = 2 p^2 + O(p^4).
\end{equation} 

\noindent
A careful analysis shows that the lowest order contribution
to $P(G=3)$ is of order $p^4$,
because at least four probabilistic edges are required. 
Therefore, to leading order
we obtain

\begin{eqnarray}
P(G=1) &=& 1 - 2 p^2 + O(p^4)
\nonumber \\
P(G=2) &=& 2 p^2 + O(p^4)
\nonumber \\
P(G=3) &=& O(p^4).
\label{eq:Pm}
\end{eqnarray}

\noindent
Truncating the distribution $P(G=g)$ at a degree $g=g_{\rm max}$,
its moments can be expressed by

\begin{equation}
\langle G^n \rangle = \sum_{g=1}^{g_{\rm max}} g^n P(G=g).
\end{equation}

\noindent
Taking $g_{\rm max} = 2$,
we find that 
$\langle G \rangle = 1 + 2 p^2$
and
$\langle G^2 \rangle = 1 + 6 p^2$.

\section{The distribution of shortest path lengths}

Consider an instance of the corded ND network 
with a 
distance matrix 
$L_t$
of dimensions
$N_t \times N_t$,
where 
$L_t(i,j)=\ell_{ij}(t)$
is the distance betwen nodes $i$ and $j$ at time $t$.
A splendid property of the corded ND model 
is that the addition of the daughter node
never shortens the distance between any pair of existing nodes,
$i$ and $j$,
in the network, namely 
$\ell_{i,j}(t)=\ell_{i,j}$
is fixed.
Thus, the distance matrix $L_{t+1}$ consists of the matrix $L_t$,
with the addition a row (and a column) which account for the distances between
the daughter node, D, and the rest of the network.
This property enables us to express the DSPL at time $t+1$ 
as a superposition 
of the DSPL at time $t$ and the DSPL between the daughter node, D, and
the rest of the network.

Choosing a random node, $i$, one can describe the shell
structure around such a node by the distance distribution

\begin{equation}
P_t(L = \ell) = \frac{N_t(L=\ell)}{N_t-1},
\label{eq:Pell_def}
\end{equation}

\noindent
where $N_t(L=\ell)$ is the number of nodes in the 
shell at distance $\ell$ from node $i$.
At each time step, $t$, 
a random node M, referred to as a mother node, is chosen for duplication.
The new, daughter node, D, is then connected to the mother node,
and with probability $p$ to each one of its neighbors.
The shell structure around the daughter node is 
closely related to that of the
mother node. 
Among the neighbors of the mother node, 
those of the neighbors for which the link to M is 
copied, end up at distance $L=1$ from D. 
Those neighbors of M for which the link 
to M is not copied end up at distance $L=2$
from D.
Therefore, the first shell around 
the daughter node is given by

\begin{equation}
P_t^{\rm D}(L=1) = 
p 
P_t^{\rm M}(L=1) 
+ \frac{1}{N_t-1},
\label{eq:Pd1}
\end{equation}

\noindent
where $P_t^{\rm M}(L=\ell)$
is the distance distribution around the mother node.
Thus, nodes which are at distance $L=\ell$ from the mother node,
may end up either at distance $L=\ell$ or at distance $L=\ell+1$ 
from the daughter node.
To exemplify this property, consider a target node T at distance $L=\ell$
from the mother node, M. 
A shortest path from M to T consists of a 
set of nodes
${\rm M},r_1,r_2,\dots,r_{\ell-1},{\rm T}$
in which subsequent nodes are connected.
In the case that the edge between M and $r_1$ is copied,
node T ends up at a distance $L=\ell$ from D,
while in case it is not copied node T ends up at a distance $L=\ell+1$
from D.
In the case that there is a single shortest path from M to T, the
former scenario would occur with probability $p$ while the latter
scenario would occur with probability $1-p$,
namely

\begin{equation}
P_t^{\rm D}(L=\ell) = p P_t^{\rm M}(L=\ell) +
(1-p) P_t^{\rm M}(L=\ell-1),
\end{equation}

\noindent
where $\ell \ge 2$.
However, since the shortest path from M to T may be degenerate,
the calculation of 
$P_t^{\rm D}(L=\ell)$ 
requires a more careful attention.
We express the DSPL between the daughter node, D, and the rest 
of the network in the form

\begin{equation}
P_t^{\rm D}(L=\ell) = \eta P_t^{\rm M}(L=\ell) + (1 - \eta) P_t^{\rm M}(L=\ell-1).
\label{eq:Pdell}
\end{equation}

\noindent
where 
$\ell \ge 2$
and
$0< \eta <1$. 
The assumption made here is that $\eta = \eta(p)$ does not depend on the path length $L$.

In order to evaluate the parameter $\eta$,
consider a random target node, T, which is at distance $\ell$ from the mother node, M.
In the simplest case, the shortest path, of length $\ell$
from M to T is unique.
However, it may be degenerate, in which case there
are several paths of length $\ell$
from M to T.
Here we are concerned with 
the degeneracy of the first step
along the shortest paths. 
This degeneracy is given by
the number of nearest 
neighbors of M which reside on at least
one shortest path from M to T,
and is denoted by
$G_{\rm MT}$.
Clearly, $G_{\rm MT} \le k_{\rm M}$, where $k_{\rm M}$ is the degree of the mother
node, M.

Consider a pair of nodes M and T, which are at a distance
$L=\ell$ from each other, 
where the degeneracy level of the shortest paths is given by $G_{\rm MT}=g$.
In the case that node M is chosen for duplication,
if none of the $g$ links of M 
which reside on shortest paths to T
are duplicated,
the distance between the daughter node 
D
and T 
becomes $L=\ell+1$, while in the case that at least
one of these $g$ edges is duplicated, the distance
is $L=\ell$.
Since each link of the mother node, M, is duplicated
with probability $p$, the probability that none of them
is duplicated is $(1-p)^g$.
The probability
that at least one of these $g$ links will be duplicated is 
$1-(1-p)^g$.
In order to account for the probabilistic nature of the degeneracy,
we denote the probability that 
the first step in the shortest path between random nodes M and T
is $g$-fold degenerate
by 
$P_t(G=g)$. 
Thus, the probability, $\eta=\eta(p)$, that
at least one of the $g$ neighbors of the mother node, M,
which reside along shortest paths  to T are 
connected to the daughter node can be expressed by

\begin{equation}
1 - \eta =  \sum_{g=1}^{\infty}  (1-p)^{g} P(G=g),
\label{eq:eta}
\end{equation}

\noindent
or more concisely by

\begin{equation}
1 - \eta =  \left\langle (1-p)^{G} \right\rangle.
\label{eq:etalr}
\end{equation}

\noindent
In fact, Eq. (\ref{eq:eta}) can also be expressed in 
the form

\begin{equation}
1 - \eta = F(1-p),
\label{eq:1-eta}
\end{equation}

\noindent
where

\begin{equation}
F(x) = \sum_{g=1}^{\infty}  x^g P(G=g)
\end{equation}

\noindent
is the generating function of $P(G=g)$.

For simplicity we assume that the distribution
$P(G_{\rm MT}=g)$ 
does not depend on $L_{\rm MT}$,
except for the case of $L_{\rm MT} = 1$,
in which
$P(G=g)=\delta_{g,1}$.
If this assumption holds, it guarantees that the 
assumption made above that $\eta$ 
is independent of $L_{\rm MT}$ is valid.

Using the binomial expansion of 
$(1-p)^g$
in Eq.
(\ref{eq:eta}),
it can be expressed in the form

\begin{equation}
\eta = - \sum_{n=1}^{\infty} (-1)^n B_n p^n,
\label{eq:etabinom}
\end{equation}

\noindent
where

\begin{equation}
B_n = \sum_{g=n}^{\infty} {
\binom{g}{n}
} P(G=g)
\end{equation}

\noindent
is the $n$th binomial moment of $P(G=g)$.
The first two terms in this expansion are
$\eta = B_1 p - B_2 p^2$,
where 
$B_1 = \langle G \rangle$
and 
$B_2 = (\langle G^2 \rangle - \langle G \rangle)/2$.
Taking the first term in
Eq. (\ref{eq:etabinom}),
where 
$B_1 = 1 + 2 p^2$,
we obtain

\begin{equation}
\eta = p + 2 p^3 + O(p^4).
\label{eq:eta2}
\end{equation}

\noindent
While paths of length $L=1$ are non-degenerate,
for simplicity we replace the parameter $p$ by $\eta$
also in the equation for $P_t^{\rm D}(L=1)$.
Since $p$ and $\eta$ differ from each other only in
order $p^3$, while $P_t(L=1)$ is quickly reduced to
order $1/N_t$, the error introduced by this approximation
is negligible.

Assuming that the mother node, M, is a typical node,
we replace the distribution 
$P_t^{\rm M}(L=\ell)$ 
by
$P_t(L=\ell)$.
As a result, Eqs.
(\ref{eq:Pd1}) 
and
(\ref{eq:Pdell})
are replaced by

\begin{equation}
P_t^{\rm D}(L=1) = 
\eta 
P_t(L=1) 
+
\frac{1}{N_t-1},
\label{eq:Pd1p}
\end{equation}

\noindent
and

\begin{equation}
P_t^{\rm D}(L=\ell) = \eta P_t(L=\ell) + (1 - \eta) P_t(L=\ell-1),
\label{eq:Pdellp}
\end{equation}

\noindent
respectively,
where $\ell \ge 2$.
After the node duplication step is completed, the DSPL
at time $t+1$ is given by

\begin{equation}
P_{t+1}(L=\ell) =
\frac{N_t-1}{N_t+1}  P_t(L=\ell) 
+
\frac{2}{N_t+1} P_t^{\rm D}(L=\ell)
- \frac{2P_t(L=\ell)}{(N_t-1)(N_t+1)},
\label{eq:P_td1}
\end{equation}

\noindent
where the third term on the right hand side
accounts for the dilution of the probability 
$P_{t+1}(L=\ell)$ 
due to the addition of the mother-daughter edge
to the network.
Subtracting $P_t(L=\ell)$ 
from both sides of Eq. (\ref{eq:P_td1}) 
and replacing the difference
on the left hand side
by a time derivative,
we obtain

\begin{equation}
\frac{d}{dt} P_t(L=\ell) =
- \frac{2}{N_t+1} P_t(L=\ell) + \frac{2}{N_t+1} P_t^{\rm D}(L=\ell)
- \frac{2P_t(L=\ell)}{(N_t-1)(N_t+1)},
\end{equation} 

\noindent
where $N_t=t+s$.
Plugging in the expressions for
$P_t^{\rm D}(L=\ell)$
from
Eqs.
(\ref{eq:Pd1p})
and 
(\ref{eq:Pdellp})
we obtain

\begin{eqnarray}
\frac{d}{dt} P_t(L=1) 
&=&
- 2\left( \frac{1-\eta}{t+s+1} \right) P_t(L=1)
+ \frac{2\left[ 1 - P_t(L=1) \right]}{(t+s-1)(t+s+1)},
\label{eq:P1}
\end{eqnarray}

\noindent
and

\begin{eqnarray}
\frac{d}{dt} P_t(L=\ell) 
&=&
- 2 \left( \frac{ 1-\eta}{t+s+1} \right) P_t(L=\ell) 
+
2 \left( \frac{1-\eta}{t+s+1} \right) P_t(L=\ell-1)
\nonumber \\
&-& \frac{2}{(t+s-1)(t+s+1)} P_t(L=\ell),
\label{eq:Pell}
\end{eqnarray}

\noindent
where $\ell \ge 2$.
The solution of Eqs.
(\ref{eq:P1})
and
(\ref{eq:Pell}),
for $s \ge 2$,
is given by

\begin{eqnarray}
P_t(L=1) 
&=&
\frac{s-1}{t+s-1}
\left( \frac{s+1}{t+s+1} \right)^{1-2\eta} 
P_0(L=1) 
\nonumber \\
&+&
\frac{2}{(1-2\eta)(t+s-1)}
\left[ 1 - \left( \frac{s+1}{t+s+1} \right)^{1-2\eta} \right],
\label{eq:P1s}
\end{eqnarray}

\noindent
and

\begin{eqnarray}
P_t(L=\ell) 
&=&
\left( \frac{s-1}{s+1} \right) \left( \frac{t+s+1}{t+s-1} \right)
\sum_{\ell^{\prime}=1}^{\rm{min}\{\ell,\Delta_0\}} 
\frac{ e^{-c_t} c_t^{\ell - \ell^{\prime}} }{ (\ell-\ell^{\prime})! } 
P_0(L=\ell^{\prime})
\nonumber \\
&+&
\frac{1}{(1-\eta) (s+1)} 
\left( \frac{t+s+1}{t+s-1} \right)
\sum_{\ell^{\prime}=0}^{\infty} 
\frac{ e^{-c_t} c_t^{\ell+\ell^{\prime}} }{ (\ell+\ell^{\prime})! }
e^{- \mu \ell^{\prime}},
\label{eq:Pells}
\end{eqnarray}

\noindent
for $\ell \ge 2$, where

\begin{equation}
c_t = 2 (1-\eta) \ln \left( \frac{t+s+1}{s+1} \right),
\label{eq:c_t}
\end{equation}

\noindent
and

\begin{equation}
\mu = \ln \left( \frac{2 -2\eta}{1-2 \eta} \right).
\label{eq:mu}
\end{equation}

\noindent
The parameter $\eta$ is given by Eq. (\ref{eq:eta}).
Note that for $\eta=1/2$, the exponent $e^{-\mu} = 0$, thus all the
terms in the second sum of Eq. 
(\ref{eq:Pells}) 
vanish except for the term 
$\ell^{\prime}=0$.
For $1/2 < \eta < 1$ it is convenient to replace the term
$e^{-\mu \ell^{\prime}}$ 
by

\begin{equation}
\left( \frac{1-2\eta}{2-2\eta} \right)^{\ell^{\prime}} 
= 
(-1)^{\ell^{\prime}}  \left| \frac{1-2\eta}{2-2\eta} \right|^{\ell^{\prime}}. 
\end{equation}

\noindent
Thus, for $\eta>1/2$ the second sum in Eq. (\ref{eq:Pells})
consists of positive terms for even values of $\ell^{\prime}$
and negative terms for odd values of $\ell^{\prime}$.

Eqs. 
(\ref{eq:P1s})
and
(\ref{eq:Pells})
provide a closed form expression for the DSPL of the corded ND network
at time $t$ for any size and degree distribution of the seed network.
The first term in each of these equations accounts for the effect of the
DSPL of the seed network, $P_0(L=\ell)$, while the second term does not
depend on the initial DSPL.
The first sum in Eq.
(\ref{eq:Pells})
is a convolution between the DSPL of the seed network and a Poisson
distribution. The second sum is a convolution between an exponential
function and a Poisson distribution.

Eq.
(\ref{eq:Pells})
can also be written in the form

\begin{eqnarray}
P_t(L=\ell) 
&=&
\left( \frac{s-1}{s+1} \right) \left( \frac{t+s+1}{t+s-1} \right)
\sum_{\ell^{\prime}=1}^{\rm{min}\{\ell,\Delta_0\}} 
\frac{ e^{-c_t} c_t^{\ell - \ell^{\prime}} }{ (\ell-\ell^{\prime})! } 
P_0(L=\ell^{\prime})
\nonumber \\
&+&
\frac{1}{(1-\eta) (s+1)} 
\left( \frac{t+s+1}{t+s-1} \right)
e^{-c_t (1-e^{-\mu})} e^{\mu \ell} 
\left[ 1 - \frac{\Gamma(\ell,c_t e^{-\mu})}{\Gamma(\ell)} \right],
\label{eq:PellsGamma}
\end{eqnarray}

\noindent
where $\Gamma(x)$ 
is the Gamma function and
$\Gamma(x,y)$
is the incomplete Gamma function.

\begin{figure}
\begin{center}
\includegraphics[width=8.0cm]{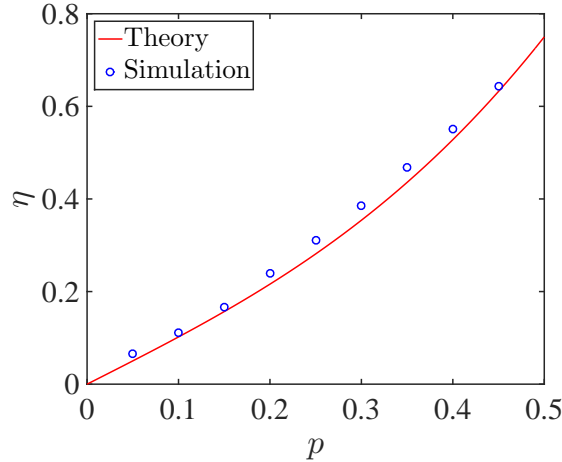}
\caption{
(Color online)
The parameter $\eta$ as a function of the probability $p$.
This parameter represents the probability 
that the distance between the daughter node, D,
and a random target node, T, is equal to the distance between the
mother node, M, and T,
namely $\eta = P(L_{DT} = L_{MT})$.
Hence,
the probability that $L_{DT} = L_{MT}+1$ is given by $1-\eta$.
The theoretical results (solid line), obtained from Eq.
(\ref{eq:eta2})
are found to be in good agreement with the simulation results
(circles).
For small values of $p$, where the shortest paths are most likely
to be unique, $\eta$ is equal to $p$. 
As $p$ is increased, the shortest paths become degenerate. 
As a result, $\eta$ acquires a nonlinear dependence on $p$,
making it larger than $p$.
}
\label{fig:4}
\end{center}
\end{figure}

In Fig. \ref{fig:4} we present the parameter $\eta$ as a
function of $p$. The theoretical results (solid line),
obtained from Eq. (\ref{eq:eta2}), are found to be in good
agreement with computer simulations (circles). 
The value of $\eta$ extracted from the simulations is the value 
which provides the best fit to the DSPL of Eq. (\ref{eq:Pells}),
when incorporated in Eq. (\ref{eq:Pdell}).
Since $\eta = \eta(p)$ increases faster than linearly with $p$,
there is a point $0 < p^{\ast} < 1/2$, for which 
$\eta(p^{\ast}) = 1/2$. 
Solving Eq. (\ref{eq:eta2}) for
$\eta(p^{\ast}) = 1/2$
we find that
$p^{\ast} =[(9 + \sqrt{105})/72]^{1/3} - 
[3 (9 + \sqrt{105})]^{-1/3}  \simeq 0.385$.

\begin{figure}
\begin{center}
\includegraphics[width=7cm]{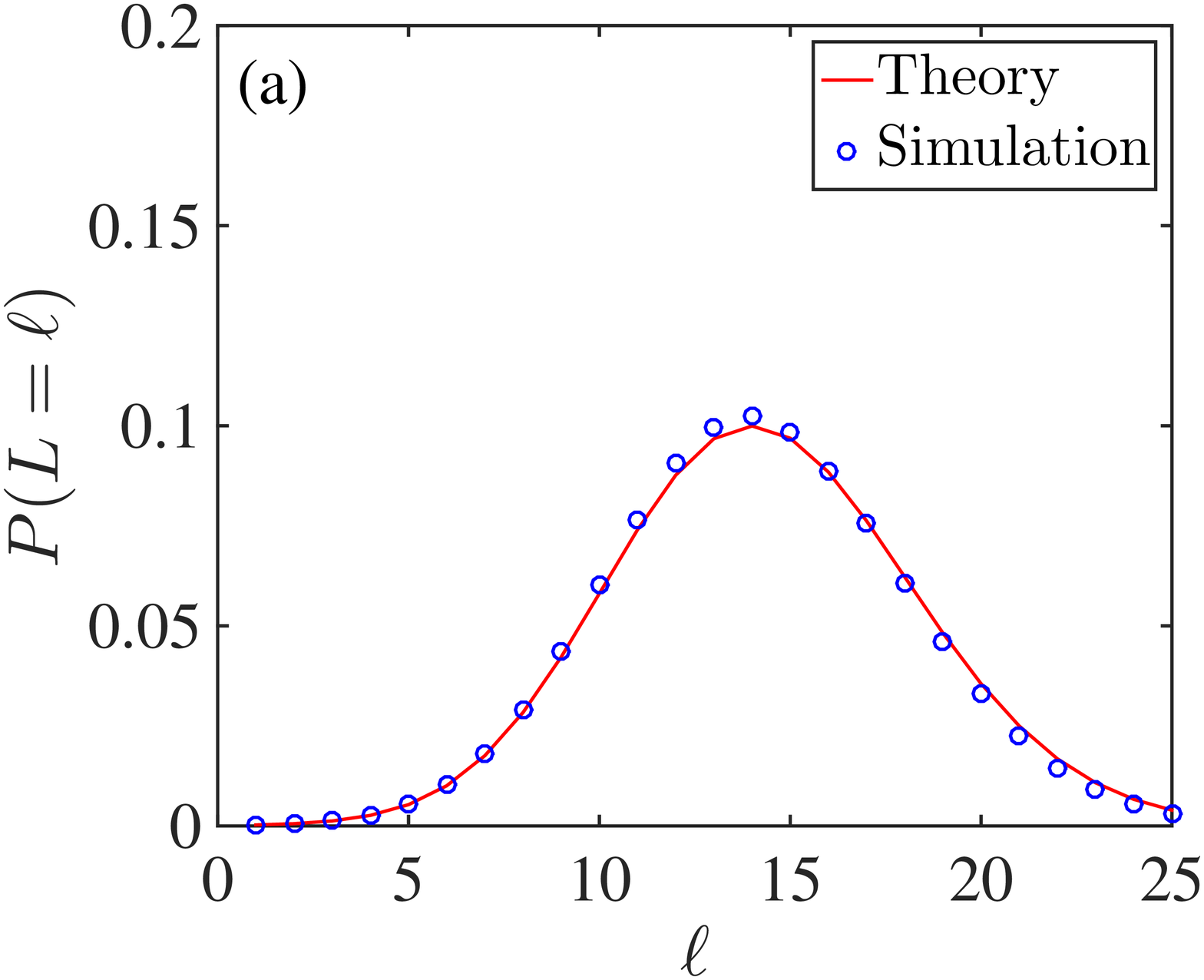} 
\includegraphics[width=7cm]{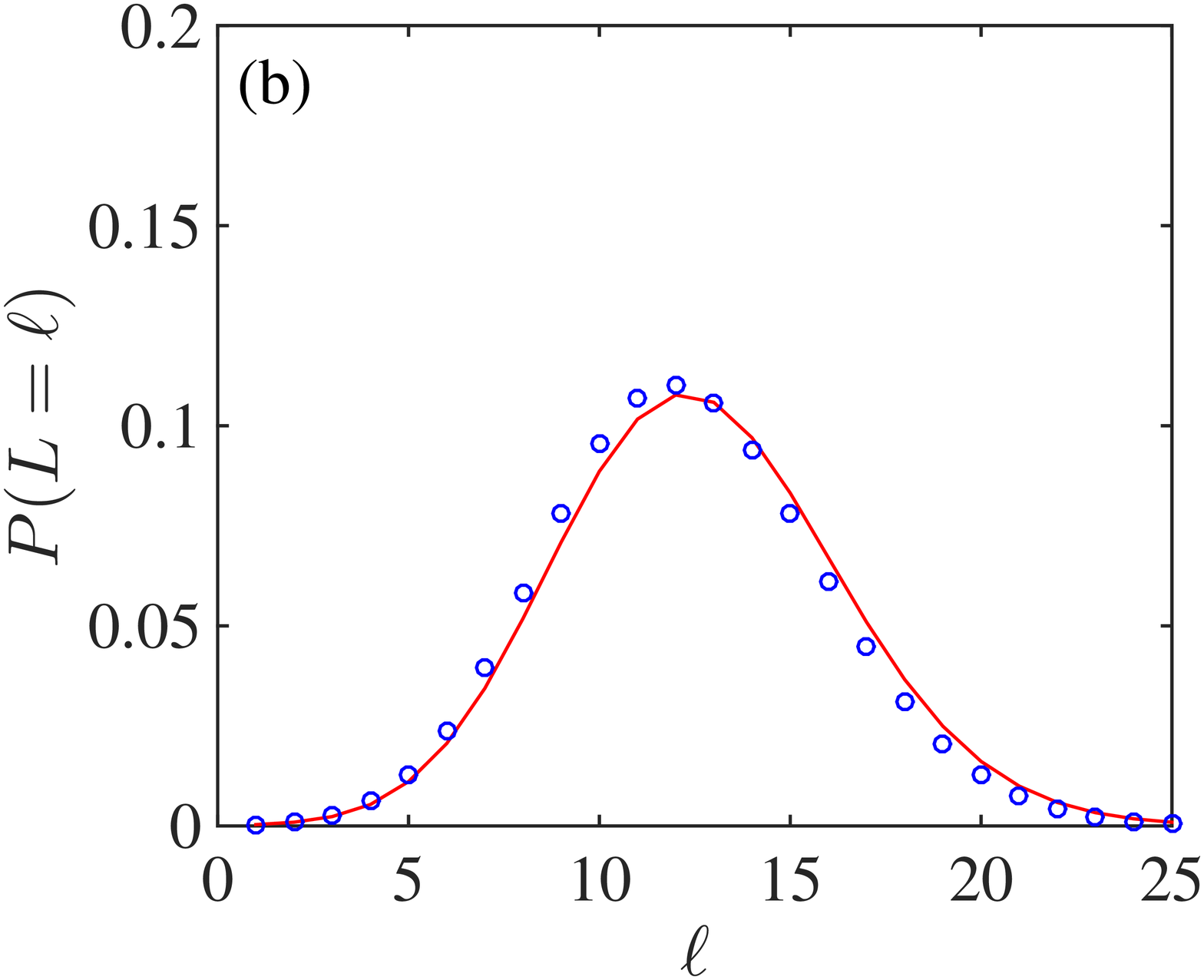} \\
\includegraphics[width=7cm]{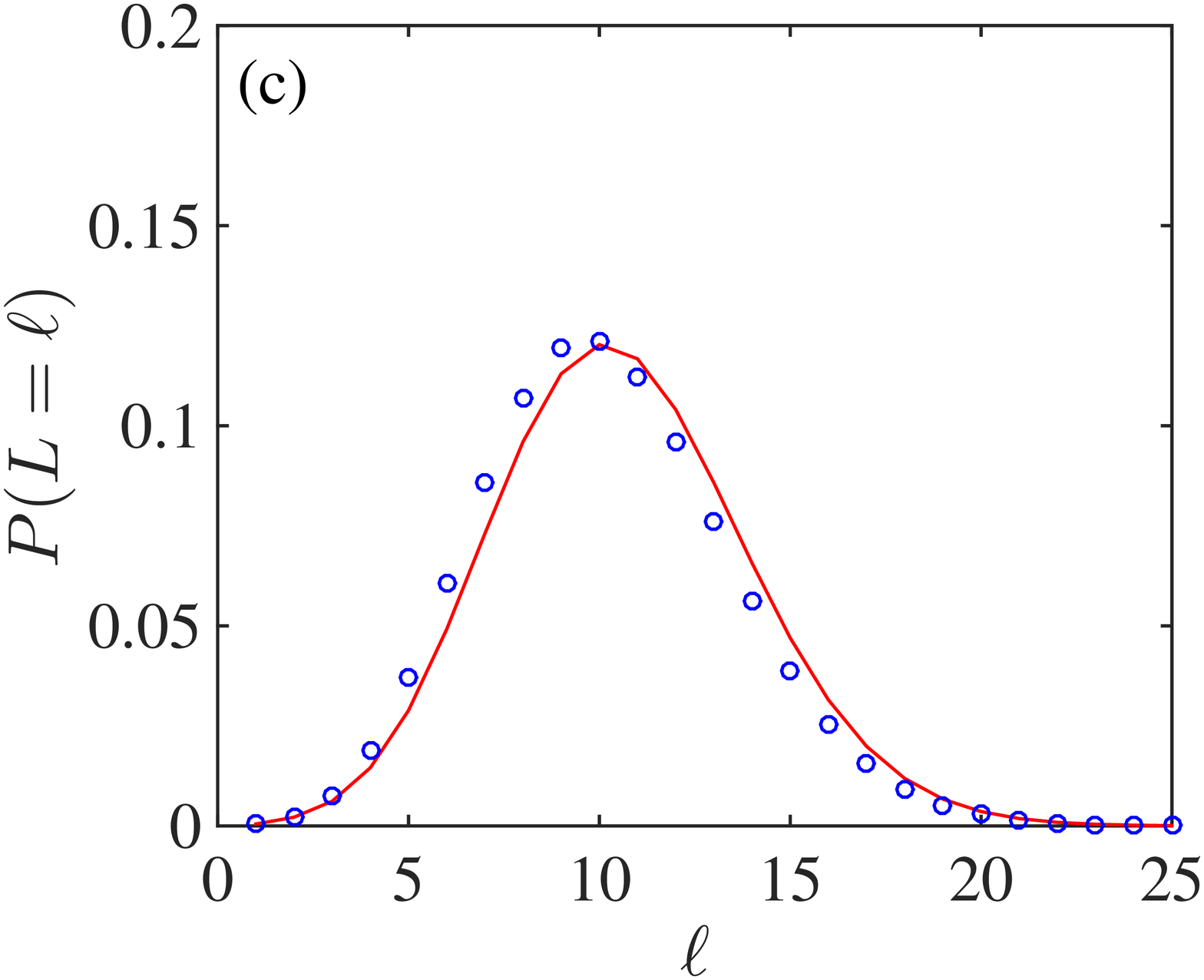} 
\includegraphics[width=7cm]{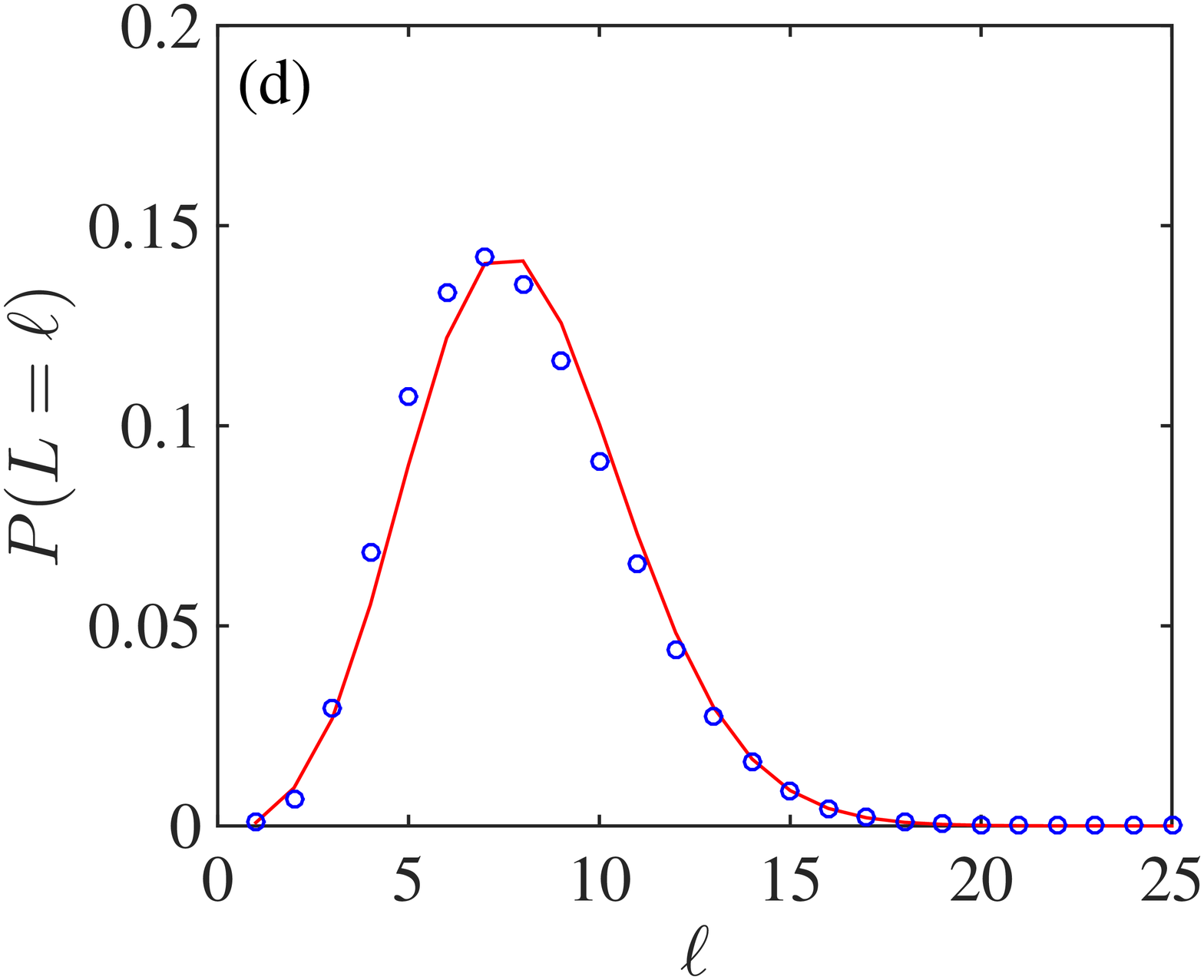} \\
\caption{
(Color online)
The DSPL of the corded ND network of $N_t=10^4$ nodes with 
(a) $p=0.1$, 
(b) $p=0.2$, 
(c) $p=0.3$,
and
(d) $p=0.4$.  
The theoretical results
(solid lines), obtained from Eqs.
(\ref{eq:P1s})
and (\ref{eq:Pells})
are found to be in good agreement with the results of computer simulations
(circles), obtained by averaging over $100$ instances.
As $p$ is increased, the distances become shorter and the DSPL becomes narrower,
consistent with Eqs. 
(\ref{eq:<d>ts2})
and
(\ref{eq:sigma2}).
The agreement is better for smaller values of $p$.
In fact, Eqs.
(\ref{eq:P1s})
and 
(\ref{eq:Pells})
are exact, while the deviation from the simulation results
are due to the underestimate of $\eta$, 
as can be seen in Fig. \ref{fig:4}.
}
\label{fig:5}
\end{center}
\end{figure}

In Fig. \ref{fig:5}
we present the DSPL, 
denoted by $P_t(L=\ell)$
vs. $\ell$ 
for an ensemble of corded ND networks of size $N_t=10^4$,
grown from a seed network of size $s=2$,
with
$p=0.1, 0.2, 0.3$ and $0.4$.
For small values of $p$, the analytical results 
(solid lines) are in very good agreement with
the simulation results (circles).
As $p$ is increased, the analytical results 
become shifted to the right compared to the
simulation results.
The simulation data was averaged over 100 network instances.

\section{Properties of the DSPL}

The first sum in Eq.
(\ref{eq:Pells})
accounts for paths which emerge from repeated duplication
of nodes and edges along paths of the seed network.
It can be noted that the probability 
$P_t(L=\ell)$ 
is affected only by the
initial probabilities 
$P_0(L=\ell^{\prime})$ 
for which 
$\ell^{\prime} \le \ell$.
This is due to the fact that the distance from a daughter node to any other
node in the network is equal or larger by $1$ than the distance from the
mother node. 
The second sum accounts for repeated duplication of nodes and edges
along new paths that emerge beyond the seed network. The exponential
function accounts for the backbone tree structure which
emerges from the edges connecting the mother and daughter nodes.
The Poisson distribution accounts for the probabilistic connections to 
the neighbors of the mother node.
Both sums in Eq.
(\ref{eq:Pells})
involve the same Poisson distribution,
$P(m) = e^{-c_t} c_t^m/m!$,
whose mean, $c_t$
is given by Eq. 
(\ref{eq:c_t}).
The first sum runs over terms in the range
$m=\ell-1,\ell-2,\dots,{\rm max} \{0,\ell-s+1\}$,
while the second sum runs over terms in the range
$m = \ell, \ell+1,\dots,\infty$.

Below we consider some special cases and limits in which
the expression for $P_t(L=\ell)$ can be simplified.
In particular, we study specific choices of the seed network,
such as a complete graph of $s$ nodes, and the special
case of a single node, in which $s=1$.
We also consider specific
values of the parameter $p$, such as $p=0$,
in which the corded ND network is reduced to the backbone tree.
Another special case is the value of $p$ for which
$\eta(p)=1/2$. 
In this case,
the parameter $\mu$ diverges.
As a result, the exponentials, 
$e^{-\mu \ell^{\prime}}$,
in the second sum
in Eq. (\ref{eq:Pells}) 
vanish, except for the term with $\ell^{\prime} = 0$,
thus the sum is reduced to a single term.

A convenient choice for the seed network is
a complete graph of $s \ge 2$ nodes.
In this case 
the initial DSPL is given by
$P_0(L=1) = 1$
and
$P_0(L \ge 2) = 0$.
The expression for the DSPL at time $t$
is simplified to

\begin{equation}
P_t(L=1) =
\frac{s-1}{t+s-1}
\left( \frac{s+1}{t+s+1} \right)^{1-2\eta}
+
\frac{2}{(1-2\eta)(t+s-1)}
\left[ 1 - \left( \frac{s+1}{t+s+1} \right)^{1-2\eta} \right],
\end{equation}

\noindent
and

\begin{equation}
P_t(L=\ell) =
\left( \frac{t+s+1}{t+s-1} \right)
\left[
\left( \frac{s-1}{s+1} \right) 
\frac{ e^{-c_t} c_t^{\ell-1} }{ (\ell-1)! } 
+
\frac{1}{(1-\eta) (s+1)} 
\sum_{\ell^{\prime} = 0}^{\infty} 
\frac{ e^{-c_t} c_t^{ \ell + \ell^{\prime} } }{ (\ell+\ell^{\prime})! }
e^{- \mu \ell^{\prime} }
\right],
\end{equation}

\noindent
for $\ell \ge 2$,
where $c_t$ is given by Eq.
(\ref{eq:c_t})
and
$\mu$ is given by
Eq. (\ref{eq:mu}).

In case that the seed network consists of a single node,
$s=1$, 
the probability 
$P_0(L=\ell) = 0$ is not defined.
However, after one time step, at $t=1$, the network consists of a pair 
of connected nodes, where $P_1(L=1)=1$ and $P_1(L \ge 2)=0$.
Thus, the ensemble of networks obtained at time $t$
for a seed network of size $s=1$ is identical to the network ensemble
obtained at time $t-1$
from a seed network of size $s=2$,
namely
$P_t(L=\ell | s=1) = P_{t-1}(L=\ell | s=2)$.
The DSPL of the resulting ND network,
for $t \ge 1$,
takes the form

\begin{equation}
P_t(L=1) =
- \left( \frac{1+2\eta}{1-2\eta} \right)
\left( \frac{3}{t+2} \right)^{1-2\eta} \frac{1}{t}
+
\left( \frac{2}{1-2\eta} \right) \frac{1}{t},
\end{equation}

\noindent
and

\begin{equation}
P_t(L=\ell) =
\frac{1}{3} \left( \frac{t+2}{t} \right)
\left[
\frac{ e^{-c_t} c_t^{\ell-1} }{(\ell-1)!}
+
\frac{1}{1-\eta} 
\sum_{\ell^{\prime} = 0}^{\infty} 
\frac{ e^{-c_t} c_t^{\ell + \ell^{\prime} } }{ (\ell+\ell^{\prime})! }
e^{ - \mu \ell^{\prime} }
\right],
\end{equation}

\noindent
for $\ell \ge 2$,
where 
$c_t$ is given by Eq.
(\ref{eq:c_t})
and
$\mu$ is given by
Eq. (\ref{eq:mu}).

In case that the parameter $p=0$, each daughter node is formed with a
single edge connecting it to its mother node.
In this case, the corded ND network is reduced to the backbone tree.
Upon formation of the daughter node, all the paths 
from it to existing nodes 
go through the mother node.
They are thus longer by $1$ than the paths starting from the mother node.
In this case, 
Eq. (\ref{eq:P1s}) 
is simplified to

\begin{equation}
P_t(L=1) 
= 
\frac{(s-1)(s+1)}{(t+s-1)(t+s+1)} P_0(L=1)
+
\frac{2t}{(t+s-1)(t+s+1)}.
\end{equation}

\noindent
In case that $p=0$ the parameters $\eta$ and $\mu$ take 
the values
$\eta=0$
and 
$\mu=\ln 2$.
Thus, Eq. (\ref{eq:PellsGamma})
is reduced to

\begin{eqnarray}
P_t(L=\ell) 
&=&
\left( \frac{s-1}{s+1} \right) \left( \frac{t+s+1}{t+s-1} \right)
\sum_{\ell^{\prime}=1}^{\rm{min}\{\ell,\Delta_0\}} 
\frac{ e^{-c_t} c_t^{\ell - \ell^{\prime}} }{ (\ell-\ell^{\prime})! } 
P_0(L=\ell^{\prime})
\nonumber \\
&+&
\left( \frac{t+s+1}{t+s-1} \right)
\left( \frac{e^{ - c_t / 2 } 2^{\ell} }{s+1} \right) 
\left[ 1 - \frac{\Gamma(\ell,c_t / 2)}{\Gamma(\ell)} \right],
\label{eq:PellsGamma2}
\end{eqnarray}

\noindent
where

\begin{equation}
c_t = 2 \ln \left( \frac{t+s+1}{s+1} \right).
\end{equation}

Another interesting case appears 
for $p = p^{\ast}$, 
where 
$\eta = \eta(p^{\ast}) = 1/2$.
In this case 
Eq. (\ref{eq:Pells})
is reduced to

\begin{equation}
P_t(L=\ell) =
\left( \frac{t+s+1}{t+s-1} \right)
\left[
\left( \frac{s-1}{s+1} \right) 
\sum_{\ell^{\prime}=1}^{{\rm min}\{\ell,\Delta_0\}} 
\frac{ e^{-c_t} c_t^{\ell - \ell^{\prime} } }{(\ell - \ell^{\prime})! } 
P_0(L=\ell^{\prime})
+
\frac{2}{s+1} 
\frac{ e^{-c_t} c_t^{\ell} }{\ell! }
\right].
\label{eq:Pell2}
\end{equation}

\noindent
For the special case in which the seed network is a complete graph,
Eq. (\ref{eq:Pell2})
is further redued to the form

\begin{equation}
P_t(L=\ell) =
\left( \frac{t+s+1}{t+s-1} \right)
\left[
\left( \frac{s-1}{s+1} \right) 
\frac{ e^{-c_t} c_t^{\ell-1} }{ (\ell-1)! } 
+
\frac{2}{ s+1 } 
\frac{ e^{-c_t} c_t^{\ell} }{\ell! }
\right],
\label{eq:Pell3}
\end{equation}

\noindent
where $\ell \ge 2$.

\section{The mean distance}

The mean distance between a random pair of nodes 
in the corded ND network 
is given by

\begin{equation}
\langle L \rangle_t = \sum_{\ell=1}^{\infty}
\ell P_t(L=\ell).
\label{eq:<d>tdef}
\end{equation}

\noindent
Taking the time derivative of
Eq. (\ref{eq:<d>tdef})
and plugging in the expressions for 
$d P_t(L=1)/d t$
from Eq. (\ref{eq:P1})
and for
$d P_t(L=\ell)/d t$
from Eq. (\ref{eq:Pell})
we obtain

\begin{eqnarray}
\frac{d}{dt} \langle L \rangle_t
&=& 
\frac{2 (\eta - 1)(t+s) - 2 \eta}{(t+s-1)(t+s+1)} 
\sum_{\ell=1}^{\infty}
\ell P_t(L=\ell)
+
\frac{2 (1-\eta)}{t+s+1} \sum_{\ell=1}^{\infty}
(\ell+1) P_t(L=\ell)
\nonumber \\
&+& 
\frac{2}{(t+s-1)(t+s+1)}.
\end{eqnarray}

\noindent
Rearranging terms we obtain

\begin{eqnarray}
\frac{d}{dt} \langle L \rangle_t 
&=&
- 
\frac{2}{(t+s-1)(t+s+1)} \langle L \rangle_t
+ 
\frac{2 (1-\eta)}{t+s+1}
+ 
\frac{2}{(t+s-1)(t+s+1)}.
\label{eq:d<d>dt}
\end{eqnarray}

\noindent
Solving Eq. (\ref{eq:d<d>dt}) we obtain

\begin{eqnarray}
\langle L \rangle_t 
&=&
2 (1-\eta) \left( \frac{t+s+1}{t+s-1} \right) 
\ln \left( \frac{t+s+1}{s+1} \right)
+
\left( \frac{s-1}{s+1} \right) 
\left( \frac{t+s+1}{t+s-1} \right) 
\langle L \rangle_0
\nonumber \\
&-&
\left( \frac{2}{s+1} \right)
\left(\frac{1}{1-2\eta} \right)
\left( \frac{t}{t+s-1} \right).
\label{eq:<d>t}
\end{eqnarray}

\noindent
In the long time limit, 
Eq. (\ref{eq:<d>t})
is reduced to

\begin{equation}
\langle L \rangle_t 
=
2 (1-\eta)  \ln \left(\frac{t+s+1}{s+1}\right)
+C_1 + C_2,
\label{eq:<d>ts}
\end{equation}

\noindent
where

\begin{equation}
C_1
=
\left( \frac{s-1}{s+1} \right) \langle L \rangle_0 
\label{eq:C1}
\end{equation}

\noindent
and

\begin{equation}
C_2
=
-
\left( \frac{2}{s+1} \right)
\left[ \frac{1-4 \eta (1-\eta) }{1-2\eta} \right].
\label{eq:C2}
\end{equation}

\noindent
The term $C_1$ accounts for the effect of the DSPL of the
seed network on
$\langle L \rangle_t$.
The term $C_2$ is a negative term which depends on $p$ and $s$.
For $0 < p < p^{\ast}$ 
(where $0 < \eta < 1/2$),
it is bounded in the range 
$-2/(s+1) < C_2 < 0$. 
For $p > p^{\ast}$
it becomes smaller than $-2/(s+1)$, thus reducing the 
mean distance $\langle L \rangle_t$.
In conclusion, 
in the long time limit
the mean distance scales logarithmically with the
network size, according to

\begin{equation}
\langle L \rangle_t 
\simeq
2 (1 - \eta)  \ln \left(\frac{t+s+1}{s+1}\right),
\label{eq:<d>ts2}
\end{equation}

\noindent
which means that the corded ND network is a small-world network.

\begin{figure}
\begin{center}
\includegraphics[width=8cm]{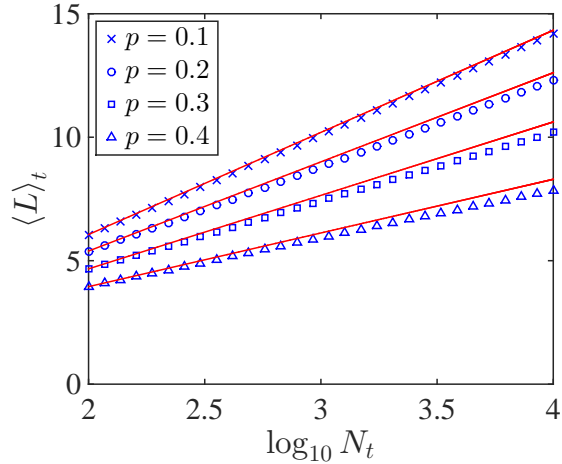}
\caption{
(Color online)
The mean shortest path length, $\langle L \rangle_t$, of the corded ND network
as a function of network size $N_t$. 
The theoretical results (solid lines), obtained from
Eq. (\ref{eq:<d>ts2}),
where $\eta$ is taken from Eq.
(\ref{eq:eta2}),
are generally in good agreement with the simulation results
(symbols),
confirming the logarithmic dependence on the network size.
As $p$ is increased, the mean shortest path length decreases.
As in Fig. \ref{fig:5}, the deviation between the theory and simulation
increases as $p$ is increased, due to the fact that  the exact value of $\eta$ is
not known.
For clarity, we focus on network sizes in the range
$10^2 \le N_t \le 10^4$.
}
\label{fig:6}
\end{center}
\end{figure}

In Fig. \ref{fig:6} we present the mean distance, $\langle L \rangle_t$,
as a function of the network size $N_t$, for $p=0.1, 0.2, 0.3$ and $0.4$.
The theoretical results, obtained from Eq. (\ref{eq:<d>ts}),
where $\eta$ is taken from Eq.
(\ref{eq:<d>ts2}), are found to be in good agreement with computer simulations (symbols).

\section{The diameter}

Consider an ensemble of corded ND networks of size $N_t$. 
In each instance of the network there are $N_t(N_t-1)/2$ pairs of 
nodes and the distances between them follow the distribution
$P_t(L=\ell)$. 
The expectation value of the number of pairs of nodes
which reside at a distance $L=\ell$ from each other is given by

\begin{equation}
N_t(L=\ell) 
=
\frac{N_t(N_t-1)}{2} P_t(L=\ell),
\end{equation}

\noindent
where $N_t=t+s$.
For sufficiently long times 
($t \gg s$), the effect of the seed network is reduced and
the DSPL exhibits a well defined peak, above which 
$P_t(L=\ell)$ 
gradually decreases.
As a result, the tail of the DSPL exhibits a 
distance $\Delta_t$,
at which
$N_t(L=\Delta_t)=1$, which can be considered as
the expectation value of the diameter of the network.
Below, we use this criterion to evaluate the diameter.
For simplicity, we consider the case in which the initial
network is a complete graph of $s$ nodes.
Note that a network resulting at time $t$ from
a seed network of size $s=1$ is equivalent to
a network at time $t-1$ with $s=2$.
Thus, in the analysis below
there is no need to treat the case of $s=1$ separately. 
Considering the large network limit, and focusing on the
large distance tail of the DSPL, it can be expressed by

\begin{equation}
P_t(L=\ell) 
=
\left( \frac{s-1}{s+1} \right)
\frac{ e^{-c_t} c_t^{\ell-1} }{(\ell-1)!}.
\label{eq:Ptt}
\end{equation}

\noindent
For convenience, we write $c_t$ in the form
$c_t = 2 (1-\eta) \ln t_s$,
where

\begin{equation}
t_s = \frac{t+s+1}{s+1}
\end{equation}

\noindent
is the network size at time $t+1$, 
expressed in units of the network size at time $t=1$.
Inserting the expression for $c_t$ into Eq.
(\ref{eq:Ptt})
and using the Stirling formula we find that

\begin{equation}
N_t(L=\ell) 
=
 \frac{(s^2 - 1) t_s^{2 \eta}}{4 (1-\eta) \ln t_s} 
\left( \frac{2 (1-\eta) e \ln t_s}{\ell} \right)^{\ell}.
\label{eq:Nt}
\end{equation}

\noindent
Inserting $N_t(L=\Delta_t)=1$
in Eq. (\ref{eq:Nt})
we obtain

\begin{equation}
\left( \frac{\Delta_t}{2 e (1-\eta) \ln t_s} \right)^{\Delta_t} 
=
\frac{(s^2-1) t_s^{2 \eta}}{4 (1-\eta)  \ln t_s}.
\label{eq:findell}
\end{equation}

\noindent
Taking a logarithm on both sides and rearranging terms,
Eq. (\ref{eq:findell})
can be expressed in the form

\begin{equation}
\left( \frac{\Delta_t}{2 (1-\eta) e \ln t} \right)
\ln \left( \frac{\Delta_t}{2  (1-\eta)e \ln t} \right)
=
\frac{
4 \eta \ln t_s 
-\ln[16 \pi (1-\eta) \ln t_s)]
+ 2 \ln \left( s^2 - 1 \right)
}
{4  (1-\eta)e \ln t_s}.
\end{equation}

\noindent
Applying the Lambert W function 
\cite{Olver2010}
on both sides
and using the relation $W(z e^z) = z$,
we obtain

\begin{equation}
\ln \left( \frac{\Delta_t}{2  (1-\eta)e \ln t_s} \right)
=
W \left[ \frac{
4 \eta \ln t_s
-\ln[16 \pi (1-\eta) \ln t_s]
+ 2 \ln \left(s^2 - 1 \right)
}
{4  (1-\eta)e \ln t_s} \right],
\end{equation}

\noindent
or

\begin{equation}
\Delta_t
= 
2  (1-\eta) 
\exp{ \left\{
1 +
W \left[ \frac{
4 \eta \ln t_s 
-  \ln \left[16 \pi (1-\eta) \ln t_s \right]
+ 2 \ln \left(s^2 - 1\right)
}
{4  (1-\eta)e \ln t_s} \right] 
\right\}}
\ln t_s
\label{eq:Delta1}
\end{equation}

\noindent
Taking the long time limit, we can approximate the argument of the $W(x)$ function.
The numerator can be replaced by its leading term, which is $2 \eta \ln t$,
thus 

\begin{equation}
\Delta_t
= 
2  (1-\eta) 
e^{1 + W \left[ \frac{\eta}{ (1-\eta)e} \right]}
\ln t_s.
\end{equation}

\noindent
Using again the above mentioned property of the $W(x)$ function,
we obtain that the expectation value, $\Delta_t$ of the diameter of the 
corded ND network is given by

\begin{equation}
\Delta_t
\simeq
\frac{2 \eta}{W \left[ \frac{\eta}{ (1-\eta)e} \right]} 
\ln \left( \frac{t+s+1}{s+1} \right).
\label{eq:Delta2}
\end{equation}

\noindent
The diameter thus scales logarithmically with the network size,
namely exhibits the same scaling as the mean distance $\langle L \rangle_t$.
However, the coefficient is larger than the coefficient of the mean distance.
Using Eqs. 
(\ref{eq:<d>ts2})
and
(\ref{eq:Delta2})
we find that 

\begin{equation}
\frac{\Delta_t}{\langle L \rangle_t} 
=
\frac{\eta}{(1-\eta) W \left[ \frac{\eta}{ (1-\eta)e} \right] }.
\label{eq:DeltaOverL}
\end{equation}

\noindent
In the dilute network limit, where $p \ll 1$, the parameter $\eta$
also satisfies $\eta \ll 1$.
Using the leading term in the Taylor expansion 
of the Lambert W function, given by
$W(x) = \sum_{n=1}^{\infty}  (-n)^{n-1} x^n / n!$,
and the relation $\eta = p + 2 p^3$, we obtain

\begin{equation}
\frac{\Delta_t}{\langle L \rangle_t} 
=
e + p + \frac{2 e -1}{2 e} p^2 + O(p^3).
\end{equation}

\noindent
Thus, in the limit of
$p \ll 1$  the diameter becomes
$\Delta_t \simeq e \langle L \rangle_t$.
This is in contrast to 
the case of configuration model networks,
where
$\Delta = \langle L \rangle + \delta$,
where $\delta$ is an additive constant
\cite{Bollobas2001,Bollobas2007}.

\begin{figure}
\begin{center}
\includegraphics[width=8cm]{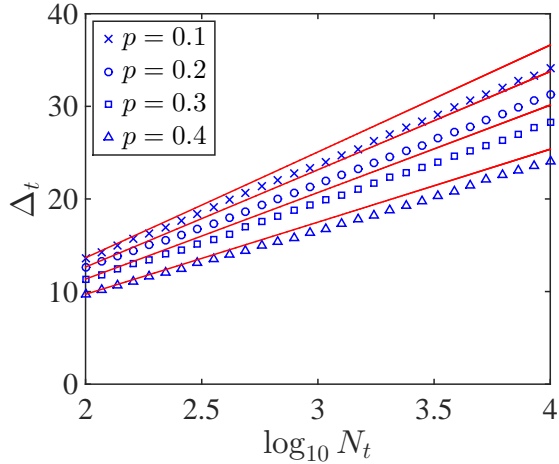}
\caption{
(Color online)
The diameter $\Delta_t$ of the corded 
ND network as a function of network size, $N_t$. 
The theoretical results (solid lines), obtained from Eq.
(\ref{eq:Delta2}),
where $\eta$ is taken from Eq.
(\ref{eq:eta2}),
are found to be in good agreement with the simulation results (symbols).
The results confirm the logarithmic dependence of the diameter on
the network size.
As $p$ is increased, the diameter decreases.
}
\label{fig:7}
\end{center}
\end{figure}

In Fig. \ref{fig:7} we present the diameter of the corded 
ND network as a function
of the network size for $p=0.1, 0.2, 0.3$ and $0.4$.
The analytical results (solid lines),
obtained from Eq.
(\ref{eq:Delta2}),
where $\eta$ is taken from Eq.
(\ref{eq:eta2}),
confirm that the diameter scales
logarithmically with the network size. 
The analytical results
over-estimate the slope compared to the simulation results (symbols).
This is due to the fact that the argument used to estimate $\Delta_t$
does not account for correlations between the longest distances in a given
instance of the network.
Thus, the result of Eq. 
(\ref{eq:Delta2}) may be considered as an upper bound for the diameter.
The simulation data was averaged over $100$ network instances.

\section{The variance of the DSPL}

In order to obtain the variance of the DSPL, 
we need to calculate its second moment, given by
$\langle L^2 \rangle_t = \sum_{\ell=1}^{\infty} \ell^2 P_t(L=\ell)$.
Taking the time derivative of 
$\langle L^2 \rangle_t$ 
and plugging in the expressions for 
$d P_t(L=1)/d t$
from Eq. (\ref{eq:P1})
and for
$d P_t(L=\ell)/d t$
from Eq. (\ref{eq:Pell})
we obtain

\begin{eqnarray}
\frac{d}{dt} \langle L^2 \rangle_t
&=& 
- \frac{2}{(t+s-1)(t+s+1)} 
\langle L^2 \rangle_t
+
\frac{4 (1-\eta)}{t+s+1} 
\langle L \rangle_t
\nonumber \\
&+&
\frac{2 (1-\eta) (t+s-1)+2}{(t+s-1)(t+s+1)},
\label{eq:d<d2>dt2}
\end{eqnarray}

\noindent
where 
$\langle L \rangle_t$
is given by
Eq. (\ref{eq:<d>ts}).
Keeping only the leading terms we obtain

\begin{equation}
\frac{d}{dt} \langle L^2 \rangle_t
= 
\frac{4 (1-\eta) [ \ln (t+s+1) +2 C_1 +2 C_2 + 1 ]}{t+s+1}. 
\label{eq:L2}
\end{equation}

\noindent
Note that 
as $p$ approaches $1/2$ from below the second term on
the right hand side becomes large and cannot be neglected in
Eq. (\ref{eq:L2}).
The solution of Eq. (\ref{eq:L2}) is

\begin{equation}
\langle L^2 \rangle_t = 
\langle L^2 \rangle_0
+
\left[ 2 (1-\eta) \ln \left( \frac{ t+s+1}{s+1} \right) \right]^2
+ 
2 (2 C_1 + 2 C_2 + 1) (1-\eta) \ln \left( \frac{t+s+1}{s+1} \right).
\end{equation}

\noindent
Thus, the variance 
$\sigma_t^2 = \langle L^2 \rangle_t - \langle L \rangle_t^2$
is given by

\begin{equation}
\sigma_t^2 = 2 (1-\eta) \ln \left( \frac{t+s+1}{s+1} \right) 
+ \langle L^2 \rangle_0 - (C_1 + C_2)^2.
\label{eq:sigma2}
\end{equation}

\noindent
In the long time limit, Eq. (\ref{eq:sigma2}) can be
simplified to the form

\begin{equation}
\sigma_t^2 = 2 (1-\eta) \ln \left( \frac{t+s+1}{s+1} \right) + O(1),
\label{eq:sigma2s}
\end{equation}

\noindent
which highlights the logarithmic scaling.
Comparing Eqs. (\ref{eq:<d>t})
and (\ref{eq:sigma2})
we find that to leading order
$\sigma_t^2 = \langle L \rangle_t$,
which is the result obtained in the case of a Poisson distribution.

\begin{figure}
\begin{center}
\includegraphics[width=8cm]{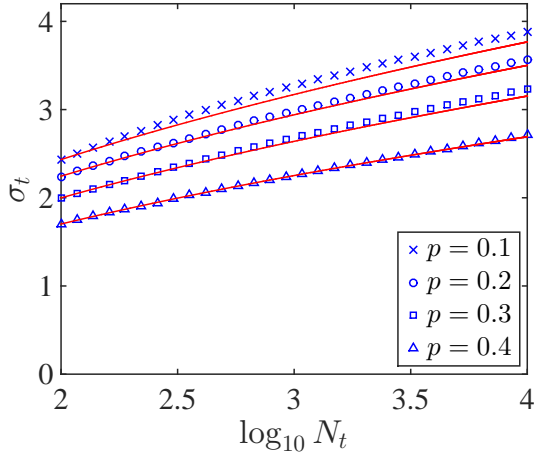}
\caption{
(Color online)
The standard deviation of the DSPL, $\sigma_t$,  
as a function of network size, $N_t$.
The theoretical results (solid lines),
obtained from 
Eq. (\ref{eq:sigma2s}),
where $\eta$ is taken from Eq.
(\ref{eq:eta2}),
are found to be in good agreement with the simulation results (symbols).
}
\label{fig:8}
\end{center}
\end{figure}

In Fig. \ref{fig:8} we present the standard deviation,
$\sigma_t$,
of the DSPL
of the corded ND model as a function of network size, $N_t$. 
The analytical results (solid lines),
obtained from Eq.
(\ref{eq:sigma2s}),
where $\eta$ is taken from Eq.
(\ref{eq:eta2}),
are found to be in good agreement with the results of numerical
simulations (symbols),
thus the logarithmic scaling is confirmed.

\section{Discussion}

The mean distance, $\langle L \rangle_t$ of the corded 
ND network was found to scale
logarithmically with the network size, $N_t$, according to
$\langle L \rangle_t \simeq 2 (1-\eta) \ln N_t$,
and it is thus a small world network.
A similar logarithmic scaling is observed in other random networks such
as configuration model networks.
However, the pre-factor of the logarithmic term is different.
In configuration model networks
the mean distance is given by 
\cite{Chung2002,Chung2003}

\begin{equation}
\langle L \rangle = 
\frac{1}{\ln \left( \frac{\langle K^2 \rangle - \langle K \rangle}
{\langle K \rangle} \right)}
\ln N.
\end{equation}

\noindent
The pre-factor of $\ln N$ is equal to the inverse of the logarithm of the
connective constant, which is expressed in terms of the first two moments
of the degree distribution.
Using Eq. (\ref{eq:etalr}),
the mean distance of the corded ND network can be expressed in the form

\begin{equation}
\langle L \rangle_t 
\simeq 
2 \left\langle (1-p)^G \right\rangle
\ln N_t.
\label{eq:LG}
\end{equation}

\noindent
Thus, the mean distance in the corded ND network is expressed in terms of the
generating function of the distribution of degeneracy levels, $P(G=g)$, 
unlike the configuration model in which it is given in terms of the first two
moments of the degree distribution, $P(K=k)$.

\begin{figure}
\begin{center}
\includegraphics[width=8cm]{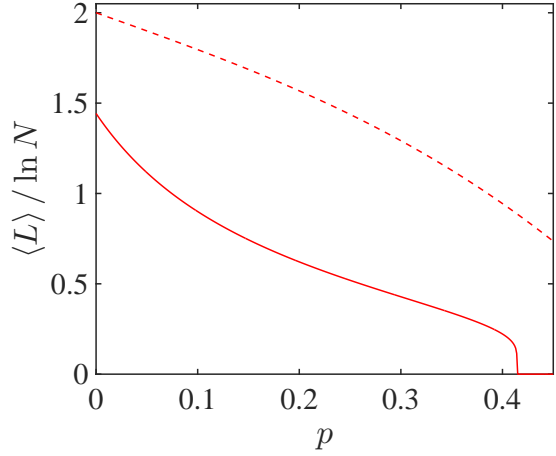}
\caption{
(Color online)
The mean distance, $\langle L \rangle = \langle L \rangle_t$,
expressed in units of $\ln N = \ln N_t$,
namely
$\langle L \rangle / \ln N \simeq 2(1-\eta)$,
of the corded ND network as a function of the parameter $p$
(dashed line),
and the corresponding ratio,
$\langle L \rangle / \ln N = 
1/\ln [(\langle K^2 \rangle - \langle K \rangle)/\langle K \rangle]$,
for a configuration model network with the same degree distribution
(solid line),
where
$\langle K \rangle$ is given by 
Eq. (\ref{eq:Kmean})
and
$\langle K^2 \rangle$
is given by
Eq. (\ref{eq:K2mean}).
}
\label{fig:9}
\end{center}
\end{figure}

In order to compare the quantitative behaviors of the corded ND network and
the configuration model network, we present in
Fig. \ref{fig:9} 
the mean distance, 
$\langle L \rangle_t$, 
expressed in units of $\ln N_t$,
of the corded ND network (dashed line)
and of the corresponding configuration model network
with the same degree distribution (solid line),
as a function of $p$.
For the corded ND network, this ratio is

\begin{equation}
\frac{\langle L \rangle_t}{\ln N_t} 
\simeq 
2(1-\eta),
\end{equation}

\noindent
where $\eta$ is given by Eq.
(\ref{eq:eta2}).
For the corresponding configuration model network, it is expressed by

\begin{equation}
\frac{\langle L \rangle}{\ln N} 
= 
\frac{1}{\ln 
\left( \frac{\langle K^2 \rangle - \langle K \rangle}{\langle K \rangle} \right)},
\end{equation}

\noindent
where
$\langle K \rangle$ is given by 
Eq. (\ref{eq:Kmean})
and
$\langle K^2 \rangle$
is given by
Eq. (\ref{eq:K2mean}).
It is found that for the corded ND network this ratio is of order 
$1$ for the whole range of sparse networks
while in the corresponding configuration model network
it decreases as $p$ is increased until
it falls sharply to zero at $p=\sqrt{2}-1$. 

We also calculated the diameter, $\Delta_t$,
of the corded ND network and found that
in the long time limit 

\begin{equation}
\frac{\Delta_t}{\ln N_t} 
\simeq
\frac{2 \eta}{W \left[ \frac{\eta}{ (1-\eta)e} \right] },
\label{eq:DeltaOverL2}
\end{equation}

\noindent
where 
$\eta = p + 2 p^3$. 
For $p \ll 1$, using the Taylor expansion 
of the Lambert W function
we obtain

\begin{equation}
\frac{\Delta_t}{\ln N_t} 
=
2 (1-\eta) \left( e + p + \frac{2 e -1}{2 e} p^2 \right) + O(p^3).
\end{equation}

\noindent
Thus, in the limit of
$p \ll 1$  the diameter becomes
$\Delta_t \simeq 2 e (1-\eta) \ln N_t$,
namely by a factor of $e$ larger than 
the mean distance, $\langle L \rangle_t$.
This is in contrast to 
the case of configuration model networks,
where
$\Delta = \langle L \rangle + \delta$,
and
$\delta$ 
is an additive constant
\cite{Bollobas2001,Bollobas2007}.

The variance of the DSPL was found to scale like

\begin{equation}
\sigma_t^2 =  2 (1-\eta) \ln N_t,
\label{eq:sigma2sd}
\end{equation}

\noindent
namely the variance scales linearly with the mean 
distance, which reflects the dominance of the Poisson 
distribution in the DSPL.
Thus, the variance of the DSPL in the corded ND network
is much larger than in the corresponding configuration model
networks, in which the DSPL tends to be narrow.

It will be interesting to generalize the analysis presented
here to the calculation of the DSPL of the 
uncorded ND network, in which there is no link between the
mother and daughter nodes.
A useful simplifying property of the 
corded ND model studied here is that
the daughter node is never discarded, namely
each randomly selected mother node
is actually duplicated.
This guarantees that the degree
of the mother node selected at time $t$ is drawn from the
instantaneous degree distribution, $P_t(K=k)$.
In the uncorded ND model this is not the case,
because the probability that the daughter node will form
a link to at least one neighbor of the mother node and thus will 
be added to the network depends on the degree of the mother node.
The conditional probability that the daughter node will be added to the network,
given that the mother node is of degree $k$, is 
$P_t({\rm added}| K=k) = 1 - (1-p)^k$.
Using Bayes' theorem, it can by shown that the 
degree distribution of the mother node
under the condition that the daughter node 
was actually added to the network is

\begin{equation}
P_t(K=k | {\rm added}) = \frac{1-(1-p)^k}{1 - G_t^0(1-p)} P_t(K=k),
\end{equation}

\noindent
where 
$G_t^0(x) = \sum_k x^k P_t(K=k)$
is the generating function of the degree distribution at time $t$.
The fact that $P_t(K=k | {\rm added})$ is different from
$P_t(K=k)$ is expected to make the calculation of the DSPL more
difficult, because the mother nodes in this case are not simply random nodes. 
The DSPL between a node, $i$, of degree $k_i$ and the rest of the network 
depends on $k_i$.
It will thus require to derive a set of master equations for 
the conditional DSPLs,
$P_t(L=\ell | K=k)$, between a random node of degree $k$ and 
all other nodes in the network.

\section{Summary}

We have studied a node duplication network model,
in which at each time step a random mother node is 
selected for duplication, referred to as the corded ND model. 
The daughter node is connected deterministically to the mother node,
and is also connected, with probability $p$, to each
one of its neighbors.
We focused on the regime of dilute networks, obtained for
$0 < p < 1/2$. 
We derived a master equation for the
time evolution of $P_t(L=\ell)$.
Finding an exact analytical solution of the master equation, we
obtained a closed form expression for the DSPL, in which
the probability $P_t(L=\ell)$
is expressed as a sum of two terms.
The first term is a convolution between the
DSPL of the seed network,
$P_0(L=\ell)$,
and a Poisson distribution.
The second term is a convolution between
a discrete exponential function 
and the Poisson distribution.
We calculated the mean distance 
$\langle L \rangle_t$
and showed that in the long time limit it scales like
$\langle L \rangle_t \simeq 2 (1-\eta) \ln N_t$,
where $N_t$ is the network size at time $t$.
The mean distance thus scales logarithmically with the 
network size, which means that the corded ND network is a small world network.
Interestingly, this behavior differs from other scale-free networks
which are ultrasmall, namely their mean distance follows
$\langle L \rangle_t \sim \ln \ln N_t$
\cite{Cohen2003}.


\begin{thebibliography}{10}


\bibitem{Albert2002}
R. Albert and A.L. Barab\'asi, 
Statistical mechanics of complex networks,
Rev. Mod. Phys. {\bf 74}, 47 (2002).

\bibitem{Caldarelli2007}
G. Caldarelli, 
{\em Scale free networks: complex webs in nature and technology} 
(Oxford University Press, 2007).

\bibitem{Havlin2010}
S. Havlin and R. Cohen,
{\it Complex Networks: Structure, Robustness and Function}
(Cambridge University Press, 2010).

\bibitem{Newman2010}
M.E.J. Newman, 
{\it Networks: an Introduction} 
(Oxford University Press, 2010).

\bibitem{Estrada2011b}
E. Estrada,
{\it The Structure of Complex Networks: Theory and Applications}
(Oxford University Press, 2011).

\bibitem{Barrat2012}
A. Barrat, M. Barth\'elemy and A. Vespignani,
Dynamical Processes on Complex Networks
(Cambridge University Press, 2012).


\bibitem{Milo2002}
R. Milo, S. Shen-Orr, S. Itzkovitz, N. Kashtan, D. Chklovskii and U. Alon,
{\it Science} {\bf 298}, 824 (2002).

\bibitem{Alon2006}
U. Alon,
{\it An Introduction to Systems Biology: Design Principles of Biological Circuits}
(Chapman and Hall/CRC, 2006).

\bibitem{Barabasi1999}
A.-L. Barabasi and R. Albert, Science {\bf 286},  509  (1999).

\bibitem{Jeong2000}
H. Jeong, B. Tombor, R. Albert, Z.N. Oltvai, and A.-L. Barab\'asi, 
{\it Nature} {\bf 407}, 651 (2000).

\bibitem{Krapivsky2000}
{P. L. Krapivsky, S. Redner and F. Leyvraz}, Phys. Rev. Lett. {\bf 85},  4629
  (2000).

\bibitem{Krapivsky2001}
P.L. Krapivsky and S. Redner, 
{\it Phys. Rev. E} {\bf 63}, 066123 (2001).

\bibitem{Vazquez2003}
A. V\'azquez, 
{\it Phys. Rev. E} {\bf 67}, 056104 (2003).

\bibitem{Milgram1967}
S. Milgram, 
{\it Psychology Today} {\bf 1}, 61 (1967).

\bibitem{Watts1998}
D. Watts and S. Strogatz,
{\it Nature} {\bf 393}, 440 (1998).

\bibitem{Chung2002}
F. Chung and L. Lu, 
{\it Proc. Nat. Acad. Sci. USA}  
{\bf 99}, 15879 (2002)

\bibitem{Chung2003}
F. Chung and L. Lu, 
{\it Internet Mathematics} {\bf 1}, 91 (2003).

\bibitem{Cohen2003}
R. Cohen and S. Havlin, 
{\it Phys. Rev. Lett.} {\bf 90}, 058701 (2003).


\bibitem{Giot2003}
L. Giot et al., 
{\it Science} {\bf 302} 1727 (2003).


\bibitem{Maayan2005}
A. Ma\'ayan, S.L. Jenkins, S. Neves, A. Hasseldine, E. Grace, B. Dubin-Thaler, 
N.J. Eungdamrong, G. Weng, P.T. Ram, J.J. Rice, A. Kershenbaum, G.A. Stolovitzky, 
R.D. Blitzer, R. Iyengar,
{\it Science} {\bf 309}, 1078 (2005).


\bibitem{Dijkstra1959}
E.W. Dijkstra,
{\it Numerische Mathematik} {\bf l}, 269 (1959).

\bibitem{Delling2009}
D. Delling, P. Sanders, D. Schultes and D. Wagner, 
Engineering Route Planning Algorithms, in 
{\it Algorithmics of Large and Complex Networks: Design, 
Analysis, and Simulation}, 
J. Lerner, D. Wagner, and K.A. Zweig (Eds.), 
p. 117 (2009).






\bibitem{Satorras2015}
R. Pastor-Satorras, C. Castellano, P. Van Mieghem
and A. Vespignani, 
{\it Rev. Mod. Phys.} {\bf 87}, 925 (2015).





\bibitem{Bollobas2001}
B. Bollobas, 
{\it Random Graphs, Second Edition}
(Academic Press, London, 2001).

\bibitem{Durrett2007}
R. Durrett,
{\it Random Graph Dynamica}
(Cambridge University Press, Cambridge, 2007).




\bibitem{Fronczak2004}
A. Fronczak, P. Fronczak, and J.A. Holyst, 
{\it Phys. Rev. E} {\bf 70}, 056110 (2004).



\bibitem{Newman2001b}
M.E.J. Newman,
{\it Proc. Natl. Acad. Sci. USA} {\bf 98}, 404 (2001).


\bibitem{Newman2001}
M.E.J. Newman, S.H. Strogatz, and D.J. Watts,
{\it Phys. Rev. E} {\bf 64}, 026118 (2001).



\bibitem{Dorogotsev2003}
S.N. Dorogotsev, J.F.F. Mendes and A.N. Samukhin, 
{\it Nuclear Physics B} {\bf 653}, 307 (2003).

\bibitem{Blondel2007}
V.D. Blondel, J.-L. Guillaume, J.M. Hendrickx and R.M. Jungers, 
{\it Phys. Rev. E} {\bf 76}, 066101 (2007).


\bibitem{Hofstad2007}
R. van der Hofstad, G. Hooghiemstra and D. Znamenski, 
{\it Electronic Journal of Probability} 
{\bf 12}, 703 (2007).

\bibitem{Esker2008}
H. van der Esker, R. van der Hofstad and G. Hooghiemstra, 
{\it J. Stat. Phys.} {\bf 133}, 169 (2008).


\bibitem{Shao2008}
J. Shao, S. V. Buldyrev, R. Cohen, M. Kitsak, S. Havlin, and H.
E. Stanley, {\it Europhys. Lett.} {\bf 84}, 48004 (2008).

\bibitem{Shao2009}
J. Shao, S. V. Buldyrev, L. A. Braunstein, S. Havlin, and H. E.
Stanley, {\it Phys. Rev. E} {\bf 80}, 036105 (2009).


\bibitem{Katzav2015}
E. Katzav, M. Nitzan, D. ben-Avraham, P.L. Krapivsky, 
R. K\"uhn, N. Ross and O. Biham,
{\it EPL} {\bf 111}, 26006 (2015).

\bibitem{Erdos1959}
P.  Erd{\H o}s and A. R\'enyi, 
{\it Publ. Math. Debrecen} {\bf 6}, 290 (1959);
{\it Publ. Math. Inst. Hungar. Acad. Sci.} 
{\bf 5}, 17 (1960);
{\it Bull. Inst. Internat. Statist} 
{\bf 38}, 343 (1961).

\bibitem{Nitzan2016}
M. Nitzan, E. Katzav, R. K\"uhn and O. Biham,
{\it Phys. Rev. E} {\bf 93}, 062309 (2016).



\bibitem{Melnik2016}
S. Melnik and J.P. Gleeson, 
arXiv:1604.05521. 


\bibitem{Molloy1995}
{M. Molloy and B. Reed},
{\it Random Struct. Algorithms} {\bf 6}, 161 (1995).



\bibitem{Bhan2002}
A. Bhan, D.J. Galas and T.G. Dewey, 
{\it Bioinformatics} {\bf 18},  1486  (2002).


\bibitem{Kim2002}
J. Kim, P.L. Krapivsky, B. Kahng and S. Redner, 
{\it Phys. Rev. E} {\bf 66}, 055101 (2002).

\bibitem{Chung2003b}
F. Chung, L. Lu, T.G. Dewey and D.J. Galas, 
{\it J. Comput. Biol.} {\bf 10}, 677 (2003).

\bibitem{Krapivsky2005}
P.L. Krapivsky and S. Redner, 
{\it Phys. Rev. E} {\bf 71}, 036118 (2005).

\bibitem{Ispolatov2005}
I. Ispolatov, P.L. Krapivsky and A. Yuryev, 
{\it Phys. Rev. E} {\bf 71}, 061911 (2005).

\bibitem{Ispolatov2005b}
I. Ispolatov, P.L. Krapivsky, I. Mazo and A. Yuryev, 
{\it New J. Phys.} {\bf 7}, 145 (2005).

\bibitem{Bebek2006}
G. Bebek,  P.  Berenbrink, C. Cooper, T. Friedetzky, J. Nadeau and S.C. Sahinalp,
{\it Theor. Comput. Sci.} {\bf 369},  239 (2006).

\bibitem{Li2013}
S. Li, K.P. Choi and T. Wu, 
{\it Theor. Comput. Sci.} {\bf 476},  94 (2013).




\bibitem{Lambiotte2016}
R. Lambiotte, P. L. Krapivsky, U. Bhat and S. Redner
{\it Phys. Rev. Lett.} {\bf 117}, 218301 (2016).


\bibitem{Bhat2016}
U. Bhat, P. L. Krapivsky, R. Lambiotte and S. Redner
{\it Phys. Rev. E.} {\bf 94}, 062302 (2016).



\bibitem{Ohno1970}
S. Ohno, 
{\it Evolution by Gene Duplication} (Springer-Verlag, New York, 1970).

\bibitem{Teichmann2004}
S.A. Teichmann and M.M. Babu, 
{\it Nature Genetics} {\bf 36},  492  (2004).


\bibitem{AutoReg}
Except for the case in which the duplicated gene is an
auto-regulator, namely a transcription factor that regulates
its own expression. In this case, one of the copies may end
up regulating the other.

\bibitem{Toivonen2009}
R. Toivonen, L. Kovanen, M. Kivel\"a, J.-P. Onnela, J. Saram\"aki and K. Kaski, 
{\it Social Networks} {\bf 31}, 240 (2009).

\bibitem{Granovetter1973}
M. Granovetter, 
{\it American Journal of Sociology} {\bf 78}, 1360 (1973).


\bibitem{Redner1998}
S. Redner,  
{\it Eur. Phys. J. B} {\bf 4}, 131 (1998).

\bibitem{Redner2005}
S. Redner,
{\it Physics Today} {\bf 58}, 49 (2005).

\bibitem{Radicchi2008}
F. Radicchi, S. Fortunato, and C. Castellano,  
{\it Proc. Natl. Acad. Sci. USA} {\bf 105}, 17268 (2008).


\bibitem{Peterson2010}
G.J. Peterson, Steve Press\'e and K.A. Dill,
{\it Proc. Natl. Acad. Sci. USA} {\bf 107 }, 16023 (2010).





\bibitem{Smythe1995}
R.T. Smythe andH. Mahmoud,
{\it Theory Probab. Math. Statist.} {\bf 51}, 1 (1995).

\bibitem{Drmota1997}
M. Drmota and B. Gittenberger,
{\it Random Struct. Alg.} {\bf 10}, 421 (1997).

\bibitem{Drmota2005}
M. Drmota and H.-K. Hwang,
{\it Adv. Appl Probab.} {\bf 37}, 321 (2005).


\bibitem{Olver2010}
F. W. J. Olver, D. M. Lozier, R. F. Boisvert, and C. W. Clark, 
{\it NIST Handbook of Mathematical Functions} 
(Cambridge University Press, Cambridge, 2010).


\bibitem{Bollobas2007}
B. Bollobas, S. Janson and O. Riordan, 
{\it Random Struct. Alg.} {\bf 31}, 3 (2007).

\end{thebibliography}
\end{document}